# Revised Response to "On the Giant Deformation and Ferroelectricity of Guanidinium Nitrate" by Marek Szafrański and Andrzej Katrusiak

Durga Prasad Karothu,[1] Rodrigo Ferreira,[1] Ghada Dushaq,[2] Ejaz Ahmed,[1] Luca Catalano,[1] Jad Mahmoud Halabi,[1] Zainab Alhaddad,[1] Ibrahim Tahir,[1] Liang Li,[1,3] Sharmarke Mohamed,[4] Mahmoud Rasras,[2] Panče Naumov[1,5]*

[1]*Smart Materials Lab, New York University Abu Dhabi, PO Box 129188, Abu Dhabi, UAE*
[2]*Division of Engineering, New York University Abu Dhabi, PO Box 129188, Abu Dhabi, UAE*
[3]*Science and Engineering Department, Sorbonne University Abu Dhabi, PO Box 38044, Abu Dhabi, UAE*
[4]*Department of Chemistry, Green Chemistry & Materials Modelling Laboratory, Khalifa University of Science and Technology, PO Box 127788, Abu Dhabi, UAE*
[5]*Molecular Design Institute, Department of Chemistry, New York University, 100 Washington Square East, New York, NY 10003, USA*

*Corresponding author. Email: pance.naumov@nyu.edu (P.N.)

Following a well-established practice of publishing commentaries to articles of other authors who work on materials that were earlier studied by them (*n.b.*, six published comments[1–6]), Marek Szafrański (MS) and Andrzej Katrusiak (AK) have recently filed on the preprint server arXiv a manuscript entitled "On the giant deformation and ferroelectricity of guanidinium nitrate"[7] with comments on our article "Exceptionally high work density of a ferroelectric dynamic organic crystal around room temperature" published in *Nature Communications* **13**, 2823 (2022).[8] Both in the submitted comment as well as in the required (by the journal) direct communication with us preceding the posting of the original comment, MS and AK have expressed dissatisfaction with the choice of literature references in our article, for which they felt that their previous work on this material has not been cited to a sufficient extent. In their original comment, they summarized their other remarks on our article as "*the structural determinations of GN [guanidinium nitrate] crystals, their phase transitions and associated giant deformation, as well as its detailed structural mechanism, the molecular dynamics and dielectric properties were reported before, while the semiconductivity, ferroelectricity, and fatigue resistance of the GN crystals cannot be confirmed.*"[7] Following our detailed response, which was submitted to the editorial office and (in a slightly longer version) filed on the same preprint server and made accessible to the public, MS and AK have now submitted a revised version of their comment, again with their established method clearly intended to discredit other authors' results as a punitive measure for not acknowledging their own. In their revised comment, they have repeatedly reiterated their dissatisfaction with the cited literature—a point that has been highlighted as many as four times—and summarized the other comments as follows: "*In conclusion, the structural determinations of GN crystals as a function of temperature and pressure, phase transitions, giant deformation, its detailed structural mechanism, molecular dynamics, and dielectric properties of GN were reported before*[2–7,9,14]." In one last attempt to address their arguments on the results advanced in our article, below, we provide a point-by-point response to MS and AK's comments. A PDF of this second, revised response at its full length will be posted on the same preprint server as the original version.

**1) On the alleged overlap of our work on guanidinium nitrate with the previous research work of MS and AK**

The work of MS and AK on guanidinium nitrate, reported three decades ago (between 1992 and 1996),[9–13] with one article published in 2004[14] and 2011,[15] focuses mainly on the structure and mechanism related to the phase transitions of guanidinium nitrate. Our research interests in this material, on the other hand, were prompted by our broader interest in dynamic crystalline materials, such as those capable of movement, shape changes, or crystals that can otherwise evolve across space and time. As with other materials that exhibit mechanical effects, **our study of guanidinium nitrate was undertaken with the specific intention of exploring the actuating performance of its crystals**. Our choice of this particular material is clearly articulated in the Introduction section of our article: "*During our broader screening for materials with thermosalient (sudden motion or shattering induced by a phase transition), actuating, and other dynamic properties we have synthesized and analyzed a number of guanidine derivatives. While most of these derivatives did not show thermosalience or other dynamic behavior, a few were found to exhibit interesting properties. The material reported here was particularly interesting for its extraordinary expansion upon heating.*"[8] Our intention to assess, compare, and rationalize the performance of the material as a dynamic actuating material is summarized in Figures 2 and 3 of our article,[8] one of which is copied as Figure 1 below, which also highlights the main conclusions of our work on this material. Our analysis, as well as most of the other results in our article, are not available from earlier studies on the material, including those reported by MS and AK. For example, our paper provides evidence for the cyclability of the mechanical response of the material; we describe the force it exerts upon transition; we report the performance indices; and we compare its performance with that of other known actuating materials and devices. Given the above-listed motivations for our work, which are clearly articulated in our article, it is disappointing to read that MS and AK have chosen to completely ignore the novelty of our work and instead incorrectly imply that our work was focused on reporting the crystal structure or detailed mechanism of the phase transition in guanidinium nitrate.

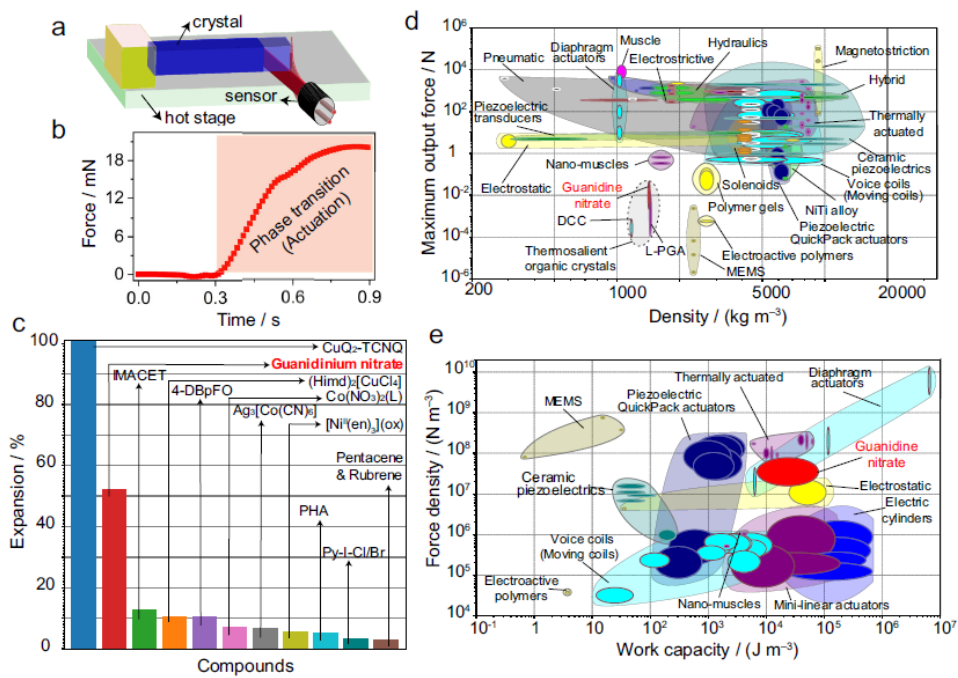

**Figure 1.** Figure 3 in our article.[8] Adapted with permission from ref. 8. Copyright 2022, Springer Nature.

In both their original and revised comments,[7] MS and AK grossly overlook the breadth of the results discussed in our paper and quite tendentiously disregard the new insights we have obtained using various techniques for structural, thermal, mechanical, electrical, and microscopic characterization that are an integral part of our report.[8] These include:

- *A modified crystallization method for the preparation of high-quality, large crystals of guanidinium nitrate.*
- *Thermal (DSC) measurements on both single crystals and lightly ground single crystals.*
- *Study of the effect of the rates of heating and cooling that has not been reported previously.*
- *Analysis of the effect of heating and cooling rates during phase transition propagation.*
- *Analysis of the thermal expansion (negative and positive thermal expansion) by using variable-temperature single crystal X-ray diffraction.*
- *Characterization of the mechanical properties of the two forms of the material by using a tensile tester at the corresponding temperatures where the phases exist.*
- *Analysis of the mechanical properties by nanoindentation.*
- *Direct microscopic measurement of the changes in the dimensions of the crystals during the phase transition on multiple crystalline samples.*
- *Characterization of the actuating performance of the crystals during the phase transition by direct measurement with a force sensor on several crystals, of which six*

- *are included in our article.[8]*
- *Comparison of the performance indices of the material with those of known existing material classes based on a large amount of literature data and visualization of the performance as Ashby plots that include the material under study.*
- *Comparison of the thermal expansion of guanidinium nitrate with other reported materials.*
- *Comparison of the expansion of the crystals of guanidinium nitrate with other reported materials.*
- *Characterization of the electrical properties of both solid-state forms (I and II) at temperatures below their transition temperatures in a four-terminal-pair configuration.*
- *Capacitance—voltage (C—V) measurements on single crystals of both forms (I and II) at temperatures below their transition temperatures.*
- *Conductance—voltage (G—V) measurements on single crystals of both forms (I and II) at temperatures below their transition temperatures.*
- *Current—voltage (I—V) measurements on single crystals of both forms (I and II) at temperatures below their transition temperatures.*
- *Frequency-dependent conductance — voltage measurements on both forms (I and II) below their transition temperatures.*
- *Dielectric measurements on single crystals of both forms at temperatures below their transition temperatures.*
- *Periodic dispersion-corrected density functional theory (DFT-D) calculations on both forms.*
- *Calculation of the energetic barrier for cell transformation in the phase transition by using CASTEP.*
- *Simulation of the cooperative rotational energy barrier by using the software VASP.*

**Comment**: *In 2022, the unprecedented strain associated with the transition between GN phases III and II was rediscovered by Karothu et al.[1]*

**Response to the comment**: In their revised comment, as well as in its original version, MS and AK unjustifiably directly attribute a "rediscovery" of the strain associated with the phase transition between forms I and II (III and II in their notation) of guanidinium nitrate to us.[7] This claim is deeply unfounded and unacceptable because **no new discovery of a "strain" or a phase transition has been explicitly stated at any point in our article**.

On the contrary, a detailed inspection of the narrative in our article would show that **our work focuses on the assessment of the performance of this material as an actuator**. The benchmarking of guanidinium nitrate as an actuator against other classes of actuating materials and devices, which is the central theme of our extensive research on this material as part of a broader ongoing study in our laboratories, is not only summarized and presented in Figures 2 and 3, but is also clearly outlined in several instances throughout the text of our article. For example, in the abstract, we state, "*We report on the exceptional performance of an organic actuating material with exceedingly large stroke...*".[8] This clearly articulates to the reader the focus of our work, which is the assessment of the *performance* of the material as an actuator (e.g., "…exceedingly large stroke…"). This, as well as other aspects of the solid-state chemistry of guanidinium nitrate, have not been investigated in the work reported by MS and AK. Moreover, in the abstract, we also write, "*This work demonstrates the hitherto untapped potential of ionic organic crystals for applications such as light-weight capacitors, dielectrics, ferroelectric tunnel junctions, and thermistors.*" In the introduction, we write, "*Here, we report that single crystals of guanidinium nitrate (GN), a ferroelectric material that undergoes rapid and reversible first-order phase transition around room temperature, <u>exhibits the highest reported stroke</u> upon transitioning between two phases due to collective reorientation of its ions in the crystal lattice.*"[8] In the summary, once again, we reiterate the goal of our research by stating that "*In summary, here we report and <u>systematically rationalize</u> the origin of the extraordinarily large thermal actuation in guanidinium nitrate crystals by comparing the crystal structures before and after the thermal phase transition.*"[8] This and other sections of the main text of our article clearly delineate our intention to quantify and compare the actuating properties of this material to those of other materials. It is therefore unclear why MS and AK insist on stating that we have claimed a (re)discovery of a strain and/or a phase transition associated with it.

## 2) On the citation of the earlier literature

**Comments:** *They cite Szafrański's paper[2] dedicated to the giant deformation of GN crystals, as their reference 48, only in the context of the GN crystallization method, while neglecting the main topic documented by the results of microscopy dilatometry and photographs of the crystal in phases III and II. No other papers[3-7,9] devoted to structural transitions of GN and its dielectric and elastic properties, that is, the main subjects of Karothu's et al. paper, were cited, either.*
............
*Moreover, previously published extensive studies on the structure, phase transitions and dielectric properties of GN crystals, i.e. the properties discussed by Karothu et al., were not acknowledged.*

**Response to the comment:** While we appreciate the feelings of MS and AK that all their previous work on GN may not have been sufficiently cited, we also note that **the choice of references in any scientific paper is ultimately at the discretion of the authors of that paper**. In our judgment, the most relevant work of MS (cited as ref. 48 in ref. 8) on this material has been duly cited in the context that was deemed appropriate. While our intention was not to purposefully neglect other previously published work by MS and AK on GN, we are reluctant to include references in our article that we consider not to be central to the focus of our paper. Contrary to the assertions of MS and AK, our attempts to provide a comprehensive list of relevant citations to prior literature actually led to a total of 80 references, which is more than the maximum recommended number of references (70) as per the guidelines of the journal. All results presented in our article have been obtained as part of our own experimental and computational work on this material and are therefore original results.

## 3) On the phase labeling and structures of phases of guanidinium nitrate

**Comment**: *In the following discussion, we will label the GN phases according to the generally accepted rule, that the highest solid phase before melting is labelled I, and the lower-temperature phases are subsequently numbered[9], i.e. differently than in reference 1. Karothu et al. redetermined the structures of GN phases III and II, without mentioning four papers reporting the same crystallographic information before[3,5,7,9]. For phase III, the* *conventional unit-cell unrelated to the layers[1], such that crystal plane (001) and unit-cell wall (**a**,**b**) lie across the layers of H-bonded ions, prevented them from connecting the deformation of the crystal at the transition to the lattice dimensions.*

**Response to the comment**: In this response to the comment, we use the notation of phases I and II used in our original article (ref. 8), which corresponds to phases III and II in ref. 7. In their comment, MS and AK have decided to enumerate the three (out of four) phases of guanidinium nitrate by using *one* of the recommended notations (III, II, I). While there is no generally accepted rule for labeling solid phases, **they do not, however, mention that in their own publications from 1993[10], 1996[12], and 1997[13], they also use the labels GN1, GN2, and GN3, i.e., nomencalture corresponding to phases I, II, and III**, which we use in our article[8] (indeed, we only use forms I and II for the two lower-temperature phases, since form III was not part of the focus of our study). Only one example of MS and AK's works with their original phase labeling is copied below:

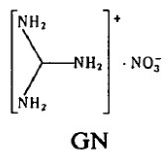

The low-temperature phase, GN1, on heating at $T_{12}$ = 296 K, undergoes a transition to phase GN2; the second phase transition occurs at $T_{23}$ = 384 K, where

the crystals transform to the GN3 phase. The transi-

**Figure 2**. Copy of the text of MS and AK's publication.[11] Adapted with permission from ref. 11. Copyright 1994, IUCr journals.

We also find it absurd the implication in the comment by MS and AK that crystal structures deposited in the Cambridge Structural Database (CSD) should not be reinvestigated by others. This stance is retrograde and restrictive to the benefits of the sophistication of modern equipment for data collection and processing, which provides data that are more precise and, in many cases, a deeper insight into the structure and properties of materials. The discussion in the comments of MS and AK implies that structural data published three decades ago should be used by others for the interpretation of properties in *all* current and future follow-up studies. Moreover, MS and AK claim that "*For phase III, the conventional unit-cell unrelated to the layers[1], such that crystal plane (001) and unit-cell wall (**a**,**b**) lie across the layers of H-bonded ions, prevented them from connecting the deformation of the crystal at the transition to the lattice dimensions*" is false; the deformation of the crystal is clearly related to the changes in lattice dimensions in Figure 2 of our original paper, which in part is copied below:

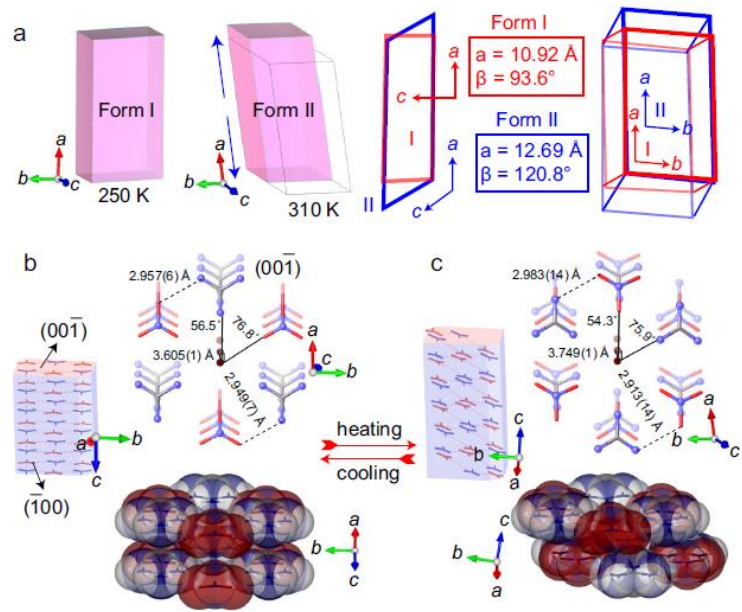

**Figure 3.** Part of Figure 2 in our article[8] that visualizes the relationship between the phase transition and crystal expansion. Adapted with permission from ref. 8 Copyright 2022, Springer Nature.

More importantly, MS and AK directly disregard the fact that, specific to our work, **the crystal structure determination of our samples was not only required but also absolutely necessary to ascertain the exact identity of the working phase** (form I or II). In our experience, the phase purity of guanidinium nitrate *critically* depends on the thermal history of the sample and/or local temperature because the phase transition occurs close to room temperature. While the existing literature on any material that has been studied in the past is definitely appreciated, at the current level of instrumental sophistication, we do not find it necessary to refer to all articles related to earlier entries in the CSD for each compound we work on, unless our purpose is actual reinvestigation of the crystal structure that has been reported previously. This is particularly the case with guanidinium nitrate, where **the structure determination in our work was performed for the purpose of phase identification**, while the previous work of MS and AK that is actually relevant to our study has already been acknowledged and cited as reference 48 of our article.

With regards to MS and AK's note on our choice of the unit cell setting[7] in their original comment, some of which is implied in the revised comment, we suspect that their phrase "unrelated to the layers" refers to the parallel molecular layers to the *ab* plane in form I. In the visualization of our structures, we have shown a cell with a *β* angle close to 90° for form I. This has only a small effect on the representation of the parallel layers from the *ab* plane in form I and does not alter the layer-wise packing or the mechanism represented in Figs. 1g, 1h, 1i, and Figures 2a, 2b, and 2c. Figure 2a shows the relation between the unit cells of forms I and II. It indicates the variation in the dimensions and shape of the unit cell and the changes in the crystallographic cell parameters (*a*, *b*, *c*, and *β*). *We have also clearly mentioned in the caption of Figure 2 that the doubling of the c axis for form II is not shown.*[8] The superimposition of the cells in Fig. 2a was intended to illustrate changes in the crystallographic cell parameters rather than to explain the mechanism of the transition. Figs. 2b and 2c show the intermolecular interactions and crystal packing along the [001] direction in forms I and II. Also, Figs. 2b and 2c show the layer-wise packing and placement of ions in adjacent layers in both forms I and II, respectively. We note that the cartooned crystal habits of forms I and II in Figs. 2b and 2c indicate the molecular orientation between adjacent layers but do not represent the actual orientation of the layers in the unit cell. This is implied by the deviation of the unit cell in form II from its actual shape (shown in Figures 2a), which could be mistaken for the other unit cell. The possible mechanism of the phase transition is clearly shown by the packing diagrams (capped stick, ball and stick, and CPK models) in Figs. 1g, 1h, 1i, and in Figs. 2b and 2c.

**4) On the quantification of the stroke**

**Comment:** *Also, the reproducibility of the large stroke associated with the first-order transition is questionable, because the GN crystals are inherently susceptible to the formation of defects.*

**Response to the comment**: In their comment,[7] MS and AK *unjustifiably question the reproducibility of our results* on the large stroke associated with the transition. However, they neglect the important fact that not only the relative expansion reported in our paper was recorded using a different method from those used in their work, but our crystalline samples were obtained by a different method than that used by them, and it cannot be expected that the two different methods and samples will yield identical results. In our work, the large crystal deformation was studied and documented from video recordings of single crystals using sophisticated optical microscopic equipment that provides direct information about the crystal expansion. As an example, we provide a copy of Figure 1e in the main text of our article:

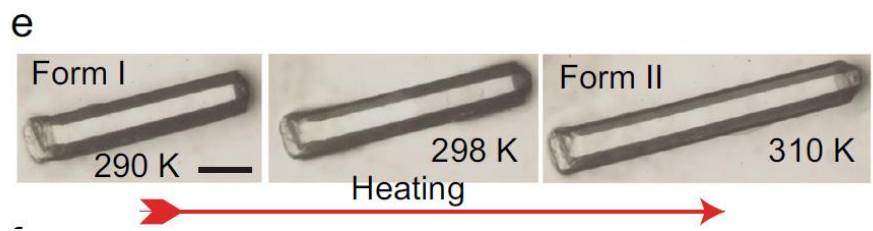

**Figure 4**. Copy of Figure 1e in our article.[8] Adapted with permission from ref. 8 Copyright 2022, Springer Nature.

These and other images in the main text and the supplementary materials of our article contrast with the monochrome photographs of poor resolution reported in an article published three decades ago by MS (cited as ref. 48 in our article)[8] copied below, where irregular crystal shape and defects are quite apparent, despite the low resolution:

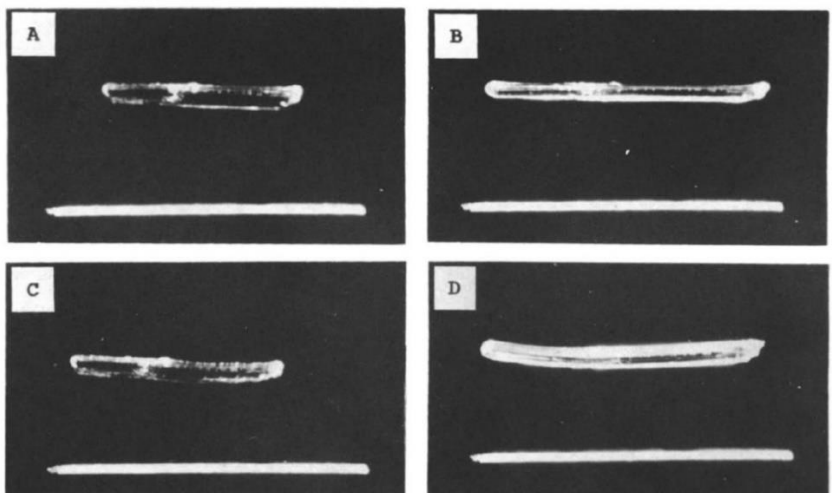

Fig. 3. Photographs of the same guanidinium nitrate crystal in both phases: (A) as grown crystal in low-temperature phase, (B) after the transition to the high-temperature phase, (C) in the low-temperature phase

**Figure 5**. Images of crystals of guanidinium nitrate published before in an article[9] that was cited in our article.[8] Note the low resolution of the images and poor quality of the crystals which is apparent from the deformed crystal shape and defects that appear as non-uniform coloration. Adapted with permission from ref. 9 Copyright 1992, Elsevier.

Unjustifiably, MS and AK go on to compare our observations and results with their own experiments ("*The deterioration of the single-crystal quality is clearly seen in the series of subsequent DSC runs presented in Supplementary Fig. 31. Three photographs presented in our Fig. 1a illustrate the typical deterioration of the crystal quality due to the generation of defects at the transition from phase II to phase III, which always reduces the deformation magnitude in subsequent cycles.*") while **quite blatantly disregarding the fact that our crystal samples and their samples have been obtained by very different methods and are of different quality**. This is even more absurd, since, in their own article from 1992,[9] they have already mentioned that the cyclability depends on crystal quality: "*It should be emphasized that the observed macroscopic changes essentially depend on the quality of the studied crystals*." Similar to their earlier samples (Figure 5),[9] Figure 1a provided in their revised comment clearly shows a number of defects in the crystal before the phase transition, some of the most prominent ones being marked with red arrows in Figure 6 below:

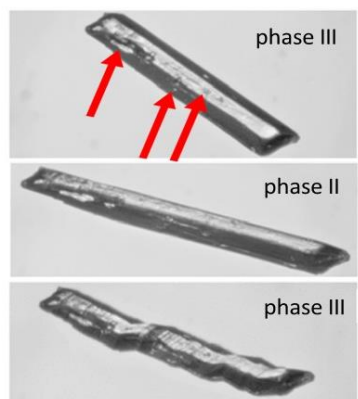

**Figure 6.** A copy of the images supplied as Figure 1a in the revised comment of MS and AK. Note the defects, some of which are highlighted with red arrows.

**Comment:** *In the following years, the crystal structure of GN phase III, built of $D_{3h}$-symmetric ions NH···O hydrogen bonded into honeycomb layers, was determined[3] and the phase transition between phases II and III was extensively characterized by calorimetric, dilatometric, powder X-ray diffraction and dielectric methods[4]. The structures of GN were studied between 4 and 395 K by X-ray and neutron diffraction as well as by neutron inelastic scattering and optical spectroscopy[5-7]. We connected the giant deformation of GN to shear lattice strain and precisely calculated from structural data the elongation of the crystal needle to 144.1% and its contraction across the needle to 68.1%[5].*

**Response to the comment:** We find strongly biased and completely unacceptable the obvious insistence of MS and AK to impose the results of their own measurement as ground truth by referring to their measurements performed by dilatometry in 1996[12] and by lattice-based calculations in 1992[9] as "precise," while they dismiss our results and the attempt to characterize the elongation directly by using microscopy and videography using modern instrumentation (for a detailed response to this comment, see below). We also disagree with MS and AK's suggestion to use their unit cell setting in order to describe the phase transition, and we refuse to accept their implicit suggestion, based on this[7] and other comments they have published[1–6] in the past, that others should refrain from experimentation and investigation of the same materials or phenomena by using alternative or complementary methods. While this contradicts common scientific research practices, it appears to be the foundation of their practice of publishing comments on others' results as a punitive measure for the lack of attention to their earlier work.[1–7]

**Comment**: *They claim the average elongation of the crystals to 151%, or even more impossible 160% (cf. their Supplementary Fig. 10K), which is incompatible with the elongation limit of 144.1% imposed by lattice strain[5]. This latter elongation is consistent with the originally reported[2] dilatometric value of 144.2%.*

**Response to the comment**: The statements of MS and AK, both in their original comment "*They claim the average elongation to 151%, or even more impossible 160% (cf. their Supplementary Fig. 10K), which is incompatible with the precise elongation to 144.1% obtained by lattice-based calculations[5] and with the dilatometric value of 144.2% originally reported[2].*" and in the revised comment, "*They claim the average elongation of the crystals to 151%, or even more impossible 160% (cf. their Supplementary Fig. 10K), which is incompatible with the elongation limit of 144.1% imposed by lattice strain[5]*", **are false**. One of the articles, cited in the original comment as ref. 5, is ref. 12, where they report "*…resulting in elongation of the sample by over 44%...*" (i.e., over 144% calculated on the total length), and an image of the original text is copied below:

reorientations of the $[C(NH_2)_3]^+$ and $NO_3^-$ ions and to a continuous phase transition at $T_{23}$. Unusually strong crystal strain accompanying the transition at $T_{12}$ and resulting in elongation of the samples by over 44% is described in terms of

**Figure 7**. A copy of the text in ref.[12] citing elongation of over 44%. Adapted with permission from ref. 12. Copyright 1996, Elsevier.

The other article cited here as ref. 2, copied below, is ref. 9. In the abstract, this article reports that "…*the length of needle-shaped crystal of guanidinium nitrate increases by about 45%.*"[9] (i.e., about 145% elongation on the total length), while in the text, it also reports "…*when the change in the crystal length at the transition to the high-temperature phase amounts to about 45%.*"

> changes in the shape of the crystals. Dilatometric measurements proved that at the transition from the low- to the high-temperature phase, the length of needle-shaped crystal of guanidinium nitrate increases by about 45%, while the transverse dimensions decrease, though the crystal is not destroyed. The phase transition is also accompanied by the changes in dielectric permittivity.

> phases, and shown in Fig. 3. Unusual character of the observed phase transition is particularly well revealed in the first cycle of heating when the change in the crystal length at the transition to the high-temperature phase amounts to about 45%. Interest-

**Figure 8**. A copy of sections of the abstract and the main text of ref.[9] stating change of crystal length of about 45%. Adapted with permission from ref. 9. Copyright 1992, Elsevier.

We do not know **why these two values, which in both original publications are given without decimal figures and with the predicators "over" and "about", are cited in MS and AK's comment with additional significant figures or how these new values were obtained**. The claimed "precise" measurements are not available from the original publications. Moreover, in their comment, MS and AK appear to have conveniently concealed the fact that in their article from 2004,[14] part of which is copied below, they report "*Our previous studies [17, 18] showed that………the former [viz. the transition between forms I and II] associated with a giant shear strain resulting in a nearly 50% elongation of the crystals,*" corresponding to an elongation of 150%, which is close to our own measured value of 151%:

> modelling a large group of functional materials. Our previous studies [17,18] showed that GN undergoes two successive phase transitions at $T_{12} = 296$ and $T_{23} = 384$ K, the former associated with a giant shear strain resulting in a nearly 50% elongation of the crystals. The

**Figure 9**. A copy of the text in ref.[14] stating "a giant shear strain" resulting in nearly 50% elongation. Adapted with permission from ref. 14. Copyright 2004, Elsevier.

As just another illustration of a biased view, both in their original and revised comments, quite tendentiously, AK and MS highlight our value of average elongation of the crystals of 151% and point out the maximum value of about 160% (sample G in the cited figure). While they make sure to mention the highest value obtained in our measurement, MS and AK do not only overlook the fact that the method used in our measurement is completely different from their method, but they also do not mention that our smallest measured value (sample J in the cited figure, copied below) is as low as 135%, a fact this reflects the naturally greater spread of values obtained with the method used in our experiment.

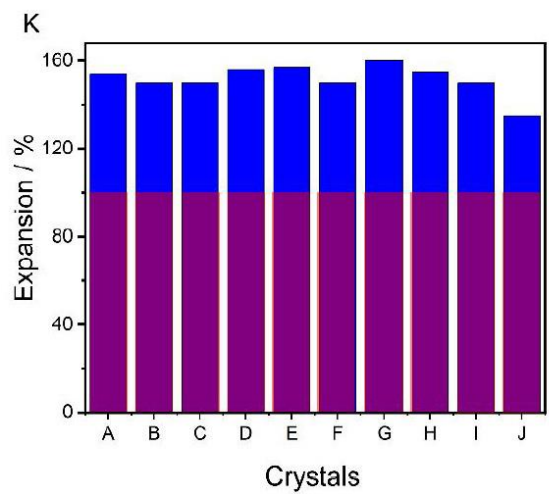

**Figure 10.** A copy of Supplementary Figure 10K in our article.[8] Our measured values vary between 135 and 160%. Adapted with permission from ref. 8. Copyright 2022, Springer Nature.

**5) On the cyclability of the phase transition between forms I and II**

**Comment**: *Karothu et al. also claim that the giant magnitude of deformation is retained through at least 20 cycles of the transitions between phases III and II without visible deterioration (their Fig. 2e), whereas even one cycle of the transition leaves structural defects (cf. their Supplementary Fig. 10), which accumulate and after very few cycles the macroscopic giant deformation disappears, as it was shown experimentally 30 years ago[2]. The deterioration of the single-crystal quality is clearly seen in the series of subsequent DSC runs presented in Supplementary Fig. 3[1]. Three photographs presented in our Fig. 1a illustrate the typical deterioration of the crystal quality due to the generation of defects at the transition from phase II to phase III, which always reduces the deformation magnitude in subsequent cycles.*

**Response to the comment**: The comment of MS and AK that "…*after very few cycles the macroscopic giant deformation disappears, as it was shown experimentally 30 years ago[2]*" **is a direct negation of our finding that guanidinium nitrate crystals of good quality can be cycled over the phase transition**. Their statement is directly and unjutifybly dismissive of our experimental evidence and biased towards their own earlier observations. Our supplementary Figure 10, copied below, shows only 10 examples of crystals *across the many crystals of varying quality that we have obtained and studied in the course of our work*. Contrary to the assertions of MS and AK, some of the crystals that we studied displayed crystal integrity without defects, while others, usually of lower quality, did develop defects. The crystals of guanidinium nitrate of good quality that we worked with did not show loss of their overall integrity on cooling, and as stated in our article,[8] they did not deteriorate visibly on heating, as we have stated on page 3 ("*Upon cooling, they shrink to their original size without any visible deterioration (Fig. 1f).*") nor upon cycling, as we have stated on page 5 ("*Indeed, the single crystals of GN are mechanically robust and as shown in Fig. 2e, they do not show evidence of fatigue even after 20 cycles, a result that reflects their extraordinary resistance to fatigue.*").

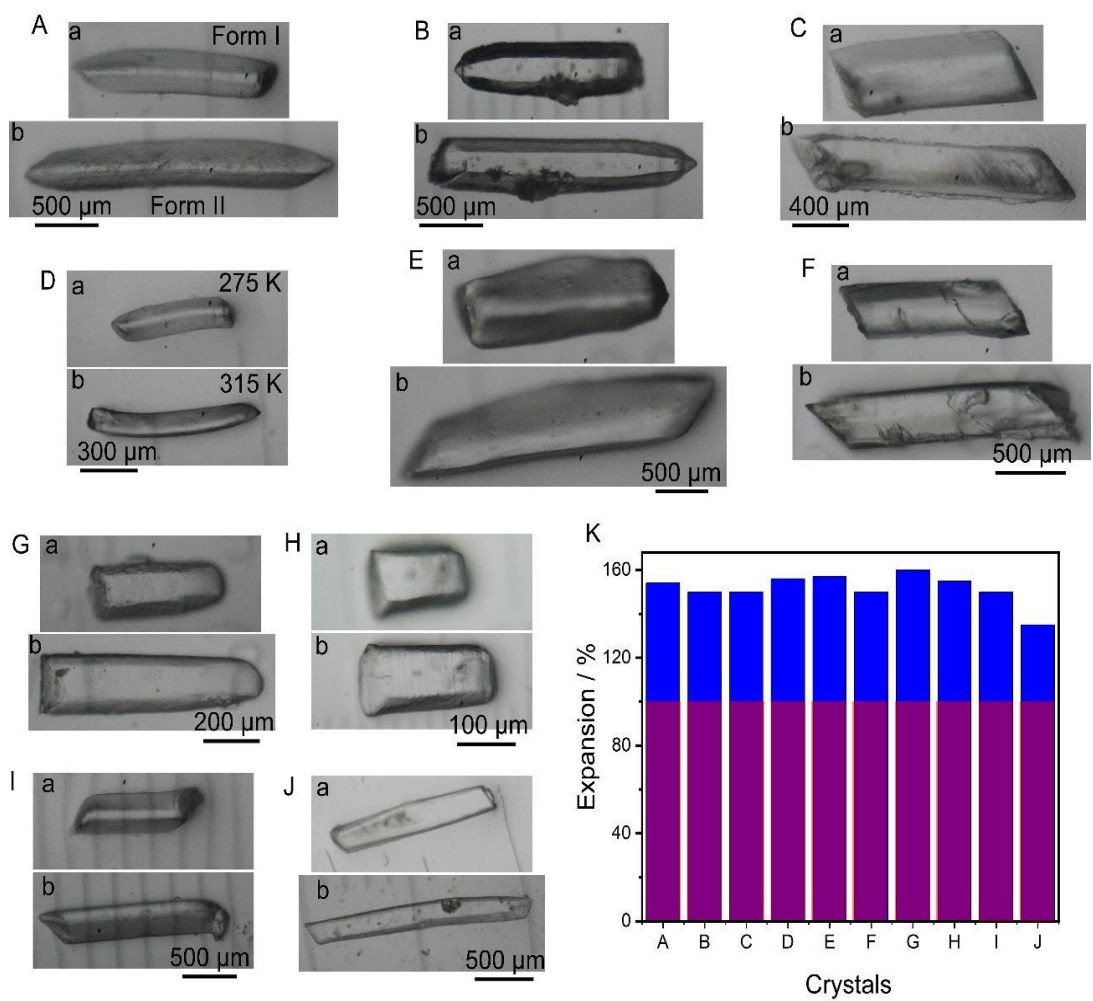

**Figure 11**. A copy of Supplementary Figure 10 in our article.[8] Adapted with permission from ref. 8. Copyright 2022, Springer Nature.

Quite tendentiously, in their statement, MS and AK refer to our Supplementary Figure 10, but they omit to mention Figure 1e (see Figure 4 above) in the main text of our paper, which shows one of the crystals of good quality that expands without deterioration. Moreover, the ability of the crystals to be cycled at least 20 times, quantified by the change in elongation of the crystal, is clearly evidenced by the results shown in Figure 2e in our article.[8] While we are not in a position to interpret or explain MS and AK's results obtained 30 years ago and their failure to observe deformation over repeated cycles, we firmly stand by our results and reaffirm our own experimental observations, which are clearly described in our article.

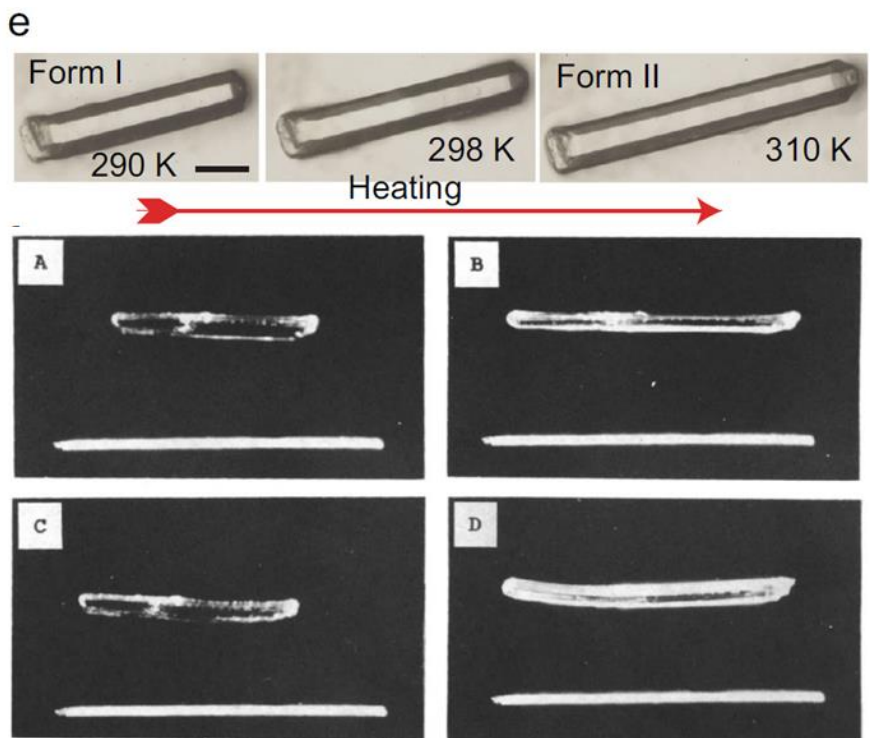

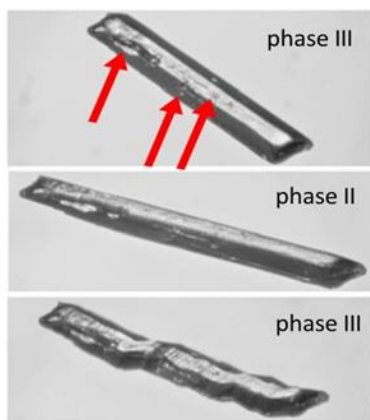

**Figure 12**. Comparison of images of crystals of guanidinium nitrate reported in our article (top),[8] in an article of MS (middle),[9] and in the revised comment by MS and AK (bottom). The red arrows highlight apparent defects in the sample before the phase transition. Adapted with permission from ref. 8. Copyright 2022, Springer Nature. Adapted with permission from ref. 9. Copyright 1992, Elsevier.

While we acknowledge that the evolution of defects is inevitable, we reitearte that **in our experience, the quality of the crystals is essential for the retention of crystal integrity**. This goes along with MS's own observation in ref. 9, copied below: "*It should be emphasized that the observed macroscopic changes essentially depend on the quality of the studied crystals.*" (Figure 13). We are not sure why in their comment MS and AK go against their own reported observations and dismiss the possibility that we may have worked with crystals of better quality.

> in Fig. 3D. It should be emphasized that the observed macroscopic changes essentially depend on the quality of the studied crystals. The crystal defects

**Figure 13**. A copy of the text published in ref.[9] Adapted with permission from ref. 9. Copyright 1992, Elsevier.

We also note that we observed that the preservation of crystal integrity, which is ultimately what accounts for cyclic operation, depends not only on crystal quality but also on the crystal shape and aspect ratio, which is why we cannot account for the failure of MS and AK to observe cyclability. In fact, we clearly state in the caption of Figure 2e, "*Cyclability of a good-quality single crystal….*" (Figure 14) and we are not sure why MS and AK disregard the statement that cyclability can be achieved only with crystals of high quality:

> unit cell parameters of form I before and during the reversible phase transition. **e** Cyclability of a good-quality single crystal over the phase transition monitored by the change in its length. The expansion and shrinkage of the crystal were recorded over 20 thermal cycles.

**Figure 14**. A copy of the caption of Figure 2 in our article.[8] Adapted with permission from ref. 8. Copyright 2022, Springer Nature.

We can only suspect that the improved cyclability observed in our experiments could be a result of our optimized method for crystallization, which might yield crystals of better quality than the ones used by MS and AK in ref. 48 (see the note on the difference in synthetic and crystallization procedures below). In their original comment, MS and AK further state that "*Even if a GN sample accidentally survived several cycles of transitions, it would not contradict the overall statistics of quickly quenched deformation, which excludes the applications requiring its repeatability,*" completely disregarding our experimental observations and confusing phenomena such as quenching in a futile attempt to disprove the prospects for potential applications of the material. Although this statement has been removed from their revised comment, we are perplexed that the authors continue to ignore others's results and experimental observations in favor of their own.

**6) On the importance of the methods for synthesis, crystallization, and handling of crystals of guanidinium nitrate**

In their comment,[7] MS and AK have grossly overlooked or intentionally decided to ignore the fact that the method for preparation of the crystals in their past work is very different from that developed by us. In their original publication,[9] they prepared guanidinium nitrate from guanidinium carbonate and nitric acid, followed by multiple crystallizations from a saturated aqueous solution below 283 K. That article[9] was cited in all their subsequent publications without providing further details on the characterization of the product after the synthesis, which may be critically important to the purity of the sample. **In our hands, the attempts to reproduce the method of MS and AK in ref. 9 did not provide good-quality crystals**, which may be the reason why they have worked with crystals of poor quality. In the revised comment, MS and AK claim to have *crystallized* guanidinium nitrate from several solvents ("*We have tested the GN crystals grown from aqueous², water-acetone¹ and water-ethanol solutions, but we detected no differences in their properties.*") without providing further experimental details. **However, in our** **experience working with this material, the solvent crystallization alone is not sufficient to prepare single crystals of good quality.** Instead, as described in the Methods section of our article, copied below,[8] we procured guanidinium nitrate from a commercial supplier (Sigma-Aldrich). Moreover, as described in our article, **we have developed *and* optimized our own protocol for purification prior to crystallization, which, in our hands, provided a large number of crystals of good quality**:

> **Crystallization.** Crystallization was based on slow evaporation of an acetone/water mixture. First, 200–400 mg of GN was taken into a conical flask. Approximately 5–10 mL of acetone was added and the conical flask was put inside the sonicator for 20–30 min at room temperature. Then, small amounts of water were added to the conical, until all the solid was dissolved. The solution was then filtered twice and the filtrate was collected in a crystallization dish, which was placed on top of ice to prevent crystallization at room temperature. Finally, the crystallization dish was taken to a crystallization room with temperature lower than 286 K at all times. GN crystals were observed in 3–4 days.

**Figure 15**. A method for preparation of crystals of good quality of guanidinium nitrate copied from our article.[8] Adapted with permission from ref. 8. Copyright 2022, Springer Nature.

In their (revised) comment, AK and MS continue to dismiss our experimental observations ("*Karothu et al. also claim that the giant magnitude of deformation is retained through at least 20 cycles of the transitions between phases III and II without visible deterioration….which accumulate and after very few cycles the macroscopic giant deformation disappears, as it was shown experimentally 30 years ago²*") and disregard our results obtained on high-quality samples, while they clearly remain biased towards their own results from nearly three decades ago. Unjustifiably, they use the results to interpret our experiments ("*The deterioration of the single-crystal quality is clearly seen in the series of subsequent DSC runs presented in Supplementary Fig. 31. Three photographs presented in our Fig. 1a illustrate the typical deterioration of the crystal quality due to the generation of defects at the transition from phase II to phase III, which always reduces the deformation magnitude in subsequent cycles.*") while **blatantly disregarding both the fact that our samples and their samples have been obtained by very different methods** and, as pointed out above, that they are clearly aware[13] that the cyclability critically depends on crystal quality (Figure 13). The short description of the preparation of their sample(s) in their revised comment indicates that the material was only *recrystallized* from different solvents, while ours were thoroughly purified prior to crystallization by a different method altogether, and moreover, the crystals were obtained and maintained below the phase transition temperature to ensure that they remain in the low-temperature phase (Figure 15). MS and AK further directly dismiss the possibility that, due to different procedures and care taken during the experiments, our samples (or at least many of our samples) may have been of higher quality than the ones used by them, and we are unsure why, in the revised version of their comment, they insist on having crystals of poor quality (Figure 5) as an argument to refute our conclusions.

It is noteworthy that in Figure 1 of their original comment, MS and AK show differential scanning calorimetry (DSC) curves, the real part of the electric permittivity, polarization-electric field dependence, and the absorption spectra of guanidinium nitrate. In their revised comment, they show images of poor-quality crystals before expansion, the same DSC curves, and the same plot of the electric permittivity as in the original comment. While it remains unclear from the caption of Figure 1 in their original comment whether the results presented were taken from their previous publications (no citation provided) or they were recorded more recently and are previously unreported results, the data in panels b and c were collected from powder samples and are therefore not directly comparable to the single crystalline samples we have reported results for. For example, the heating DSC curve shown in their original Figure 1c (Figure 1b in the revised version), copied below, shows a clearly visible shoulder on the high-temperature side of the peak around 300 K, which could indicate phase impurities introduced during synthesis, crystallization, or powdering that were not accounted for in the other measurements that also used powder samples. **Moroever, the samples measurement must have been started from room temperature. This means that the heating-cooling curve shown in Figure 16 must have been recorded after the original sample has been taken over one (if it was originally in form II) or both (if it was originally in form I) phase transitions.** That initial part of the curve is not shown, although it could be important if the effects of the history of the sample is taken into account.

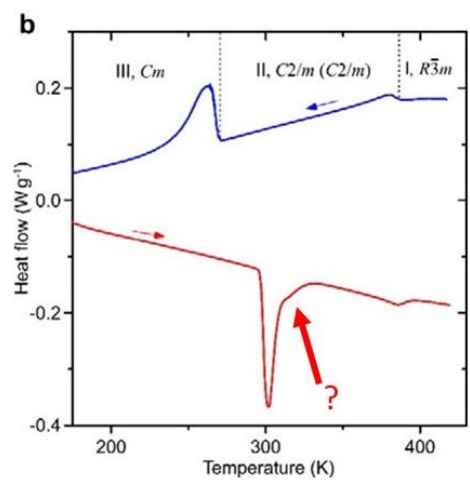

**Figure 16**. A copy of Figure 1b from the revised comment of MS and AK. The shoulder higlighted by the arrow might indicate chemical or phase impurity.

For the permittivity measurements, the authors used pressed tablets; however, **they have not confirmed the phase identity of the resulting tablets after applying pressure**, although the pressure may have altered the phase identity of the sample. No details on the preparation of the sample have been provided in Figure 1, aside from the cooling and heating rates (panel b) and a note that pressed tablets were used as samples (panel c) (the labels refer to the figure in the revised comment). Contrary to their data availability statement ("*All data obtained and analysed during this study are included in the article.*"), no details of the synthesis were provided for the samples used in these experiments, even though, in our extensive experience with this compound, **the method for preparation and handling are critically important for the outcome of the experiments for characterization**. Since there are no data that would provide *direct* evidence of the phase purity of these samples (for example, by X-ray diffraction analysis), we are unable to conclude whether they contain one or more phases or, possibly, other impurities.

In contrast with the results of MS and AK, the DSC curves reported in Figure 1d of our original article[8] were recorded from one single crystal in form I whose structure has been confirmed, and therefore the sample is devoid of phase impurities. In our experience, the transfer of the crystals to the DSC pan and the temperature of the pan are crucial to maintaining phase purity. No details are provided by MS and AK on their method for powdering the sample, although we have noted that, with a phase transition occurring around room temperature, the process of grinding guanidinium nitrate is critically important to ensure the phase purity of the sample. Their earlier characterization using X-ray powder diffraction presented in Figure 3 in their Ref.[10], which is copied below, is incorrect and confusing; the labels of panels a and b refer to GN2 and GN3, while the figure caption states GN1 and GN2, respectively. Aside from the problem with incorrect labeling, based on our own inspection, it is very probable that at least one of these samples is not a pure phase.

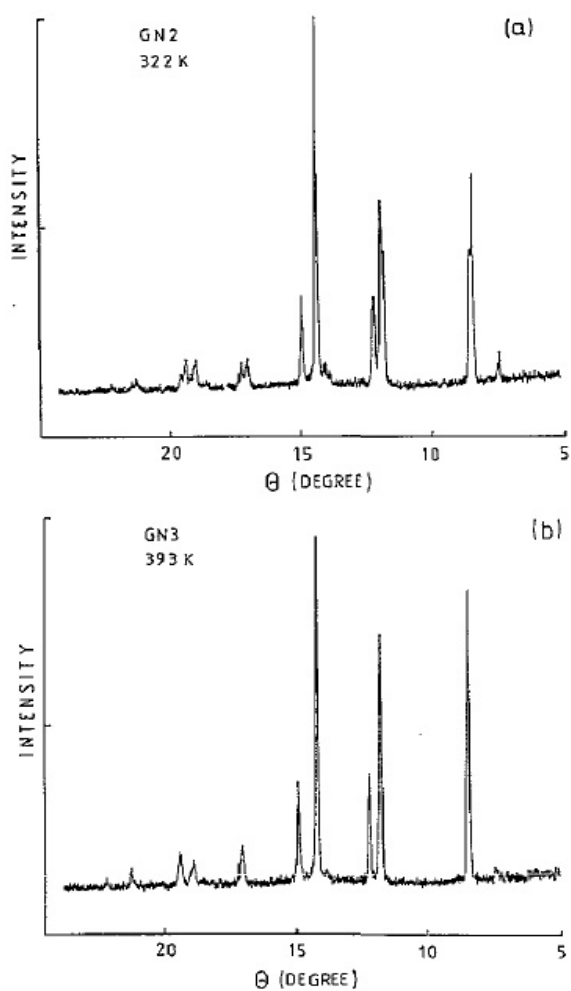

Figure 3. Diffraction patterns of the GN crystal for the phases (a) GN1 and (b) GN2.

**Figure 17**. A copy of Figure 3 from ref.[10] where the phase identity is not clear due to mislabelling of the phases in the plots and the caption. Adapted with permission from ref. 10 Copyright 1993, IOP Publishing Ltd.

**7) On the "semiconducting properties" of guanidinium nitrate**

**Comment:** *Undoubtedly, this prompted the authors to compare GN crystals with inorganic semiconductors, which is incompatible with the colourless crystal of the absorption edge in the UV region and the energy gap of 5.13 eV (see our Fig. 2d).*
*………*
*However, no generally accepted evidence of ferroelectricity and semiconductivity of this hybrid organic-inorganic material was presented.*

**Response to the comment:** We find it quite absurd and completely unacceptable that on several occasions in their original comment, that are also highlighted above as copied from their revised comment, AK and MS continue to insist that we have claimed semiconducting properties of the material under study ("*Most recently, Karothu et al.[1] described guanidinium nitrate, $[C(NH_2)_3]^+[NO^3]^-$, hereafter GN, as a ferroelectric semiconducting organic crystal with exceptional actuating properties*", "*However, no generally accepted evidence of ferroelectricity and semiconductivity of this hybrid organic-inorganic material was presented*"). In Figure 1d of their revised comment, they also provide absorption spectra to prove this point.

In our previous response, which is available online on a preprint server, we have already clearly explained that **semiconductivity as a property of guanidinium nitrate has never been studied, claimed by us, or described in our article**. Yet, MS and AK continue to argue that we have claimed that the material is a semiconductor, which is simply an incorrect and ethically unacceptable statement. In fact, a careful reading of our article[8] would confirm that no semiconducting properties of the material have been reported by us. In our article, the term "semiconductive" is mentioned only twice. In the first instance (page 6), we use the term to describe the *configuration of the experimental setup* that is usually used for such measurements; hence, the term is given in parentheses, "*(metal-semiconductor-metal configuration)*". In the second instance (page 7), we note that the response of the material *resembles* that of inorganic semiconductors. "*Interestingly, the frequency dependence of the C—V and G—V characteristics for both crystal phases are very similar to inorganic semiconductors where they exhibit…*" We are perplexed to see that these two references to technical details that were originally intended to explain better the way the experiments were performed have been misconstrued as claims of semiconducting properties of the material, which is nowhere claimed in our article. It is also unclear why MS and AK have decided to directly attribute to us explicit claims that are neither contained within our report nor have ever been the subject of our research. Here, we once again reiterate that **in our article, we do not make claims about the material being semiconductive.**

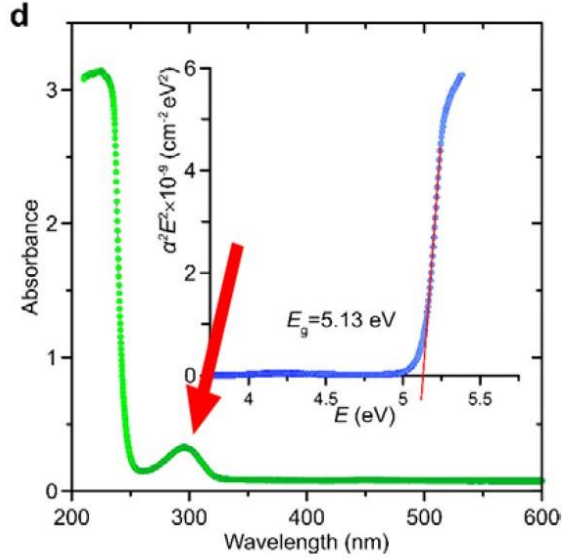

**Figure 18**. A copy of Figure 2d in the revised comment of MS and AK. The arrow marks a peak that was not discussed and assigned by the authors.

We also note that, in their attempt to prove that guanidinium nitrate is not a semiconductor, MS and AK show[7] (Figure 2d in the revised comment) an optical absorption spectrum where a clear absorption peak around 320 nm (3.87 eV) is observed, which does not correspond to the bandgap of the material at 5.13 eV. Although this has not been a focus of our study, we note that this result, which they have not commented on and which could potentially be an actual property of the material or related to phase or another impurity in their samples (see the discussion on the necessity for phase identification of the actual sample discussed above), can hardly reflect that the material is an insulator, given the difference between the optical and electrical bandgap in organic materials. It is very well established that the *color* of the crystals can reflect their optical characteristics but not necessarily their transport properties, as is apparently the case with common transparent conductive materials that are utilized for various light-related applications, such as those in solar cells. Altogether, we once again highlight that the experimental conditions in the electric characterization, type of material used, methods of preparation, and methodology used for testing the electrical performance described in the comments of MS and AK[7] are very different from those used and reported in our article.[8] Therefore, a one-to-one comparison is not only impossible but it is also unjustified.

**8) On the ferroelectric properties of guanidinium nitrate**

**Comment:** *Karothu et al. claim that GN is ferroelectric, but they fail to demonstrate the basic ferroelectric properties, like the spontaneous polarization and its switching by an external electric field, or the characteristic temperature dependence of electric permittivity in the paraelectric phase, fulfilling the Curie-Weiss law, or a ferroelectric domain structure below the Curie point[10]. The capacitance-voltage C(V) curves presented in their Fig. 4 are insufficient to prove ferroelectricity. Moreover, a typical C(V) dependence shows a maximum, corresponding to the ferroelectric polarisation switched over in the coercive field ($E_C$)[11,12]. This is due to the $\varepsilon = (1/\varepsilon_0)(\partial P/\partial E)$ relation, showing that the electric permittivity peaks when the* *polarization is reversed, hence the capacitance should be maximal at Ec. Instead, Karothu's et al. observed C(V) plots with the minima, which was not explained nor commented. When hypothetically assuming that these minima appear at the coercive field, the corresponding values of Ec in phase III would be 10–39 V/cm and in phase II 7–29 V/cm. These values could be estimated only, because the dimensions of the sample were not given[1]; on our request, the Authors responded that their samples were 0.1–0.4 mm thick. The Ec values estimated for this thickness are extremely low compared to those of true ferroelectric materials. For example, in guanidinium tetrafluoroborate and guanidinium perchlorate, built from analogous to GN honeycomb N–H···F and N–H···O bonded layers, Ec is 11.4 and 8.7 kV/cm, respectively[13]. The coercive fields of guanidinium aluminum sulfate hexahydrate (1–3 kV/cm), BaTiO$_3$ (0.5–2 kV/cm) and triglycine sulfate (0.43 kV/cm)[10] are slightly lower, but they are still 10 to 100 times higher than the values derived from the measurements by Karothu et al. The atypical shape of C(V) plots and unrealistic values of Ec prompted us to verify the discovery of ferroelectricity in GN. We have tested the GN crystals grown from aqueous[2], water-acetone[1] and water-ethanol solutions, but we detected no differences in their properties.*

**Response to the comment:** In our electrical characterization, we performed comparative testing of the electrical properties of single crystals of guanidinium nitrate in forms I and II. These were performed by using frequency-dependent capacitance-voltage (C–V), frequency-dependent conductance-voltage (G–V), and DC current-voltage (I–V) measurements. The configuration of each measurement is explained in Figs. 4a, 4b, and 5a of our article.[8] The crystal dimensions and the experimental conditions are available in the method section of the article and in the Supporting Information, and upon request, additional experimental information was made available to MS and AK.

There are many ways to test the spontaneous electric polarization and domain switching of materials, with the frequency-dependent capacitance (C) – voltage (V) hysteresis characteristic being one of the most well-known methods to study the behavior of ferroelectric capacitors.[16,17] Our measurements were performed on a parallel plate capacitor made by depositing silver on the crystal's top and bottom surfaces and sandwiching the crystal between two silver thin films. The crystal was then placed on a copper plate to access the bottom electrode, as illustrated in the schematic provided in Figure 4b of our article.[8] It should be noted that all measurements were performed on single crystals of guanidinium nitrate in form I or form II, and the phase identity of the polymorph used in the measurements was independently confirmed. Upon sweeping the voltage between -1.5 V and 1.5 V (see Figures 4e and 4f in ref. 8) at 200 kHz, a butterfly-shaped C–V curve was obtained for both crystal forms, indicating polarization switching when the direction of the alternating electric field is reversed. We note that such C–V hysteresis has been observed in other organic materials and used to confirm their spontaneous electric polarization, which is caused by polarization orientation under the applied electric field, where the electric field (E) is equal to the applied voltage (V)/(d) and d is the crystal thickness.[16–18] The C–V hysteresis is linked to the P–E loop, where the ferroelectric capacitance depends on the derivative of its polarization hysteresis.

Typically, the observed hysteresis in the capacitance-voltage (C–V) characteristics arises as a combined consequence of the inherent spontaneous polarization and the presence of interface states between the material (in this case, a guanidinium nitrate crystal) and the silver electrodes.[19] The hysteresis response, which includes aspects such as orientation and memory window, is profoundly affected by the interface states, which give rise to depolarization fields, charge injection, and electronic trapping phenomena. Consequently, the impact of these interface states can overshadow the effects attributed to polarization reversal and significantly influence polarization behavior. Specifically, the orientation of the *C–V* hysteresis can be elucidated by the existence of localized defect states present at the crystal/Ag electrode interface. This aspect has already been addressed in our article:[8] "*The four types of polarization are electronic, ionic, dipolar, and **interfacial**, and all of them can effectively add to the capacitance and conductance values in the low-frequency range.* **At high frequencies, the interfacial, dipolar, and ionic polarization contributions become insignificant, leaving only the contribution from the electronic polarization relevant.** *This frequency dependence can be clearly seen in both phases of the crystal. As shown in Supplementary Table 6, the capacitance/conductance decreases as the frequency increases from 1 to 200 kHz. This is due to the fact that at higher frequencies the temporal response of the **interfacial dipoles** is reduced, leading to a lack of rearrangement in the alternating fields direction.*" Furthermore, we write: "*Additionally, the quality of the interface between the silver and the crystal can contribute to the observed values, most probably by physical adsorption of a thin film of water on the crystal surface which can occur with testing in air.*"

Because in any given material, a number of these polarization conditions will be active at different frequencies, further investigations by *C–V* and *G–V* measurements at frequencies above 200 kHz (500 kHz – 1 MHz) on a crystal of form I were carried out by us, as reported in our original article[8]. A capacitance value of ~46 pF at 1 MHz was measured. It is widely accepted that the strength of polarization varies based on the nature of the material, the structure of its solid phase, the method of growth, and the direction of the applied field. This is clearly evident from MS and AK's responses, where they provide a range in coercive field values for different ferroelectric materials. In addition to being dismissive of the results and explanation, we also deem it unreasonable and entirely unacceptable that in their original as well as the revised response, MS and AK persist in disregarding the details of our experimental method, where the crystal dimension and the experimental conditions have been provided and are available in the method section of the article and in the Supporting Information.

**Comment:** *Our Fig. 1b shows that, apart from the first-order transition associated with the huge deformation between phases III and II, GN also undergoes a second-order transition at about 384 K. This is of primary importance, because the high-temperature phase I is centrosymmetric, of space group R3-m7. If phase III were ferroelectric, it would follow that one of these two observed phase transitions, either III-to-II or II-to-I, should be of the ferroelectric-paraelectric type. Such a transition type is accompanied by a large anomaly in the electric permittivity and therefore can be easily identified by dielectric measurements, even for polycrystalline samples. Due to the very weak frequency dispersion of the real part of electric permittivity □', we have plotted only the data for 200 kHz in Fig. 1c. Both transitions in GN are reflected in the dielectric response of the crystal, but none of them shows any features of the ferroelectric-paraelectric transition. The key test of ferroelectric properties is the spontaneous polarization switching, so we performed it both for a thin polycrystalline film and a single crystal. The linear polarization–electric field dependence and absence of ferroelectric hysteresis loop for the amplitude of the ac electric-field intensity as high as 40 kVcm$^{-1}$, clearly seen in Fig. 2a, testify that GN is non-ferroelectric. Furthermore, our C(V) tests for the single crystal of GN* *in phases III and II (Fig. 2b) show that the capacitance value is independent of the electric field, without any butterfly-shaped anomalies. Thus, all of our comprehensive studies detect no traces of ferroelectricity of GN, whatsoever.*

**Response to the comment:** In their comment, MS and AK refer to a purported second-order phase transition of guanidinium nitrate (from form II to form III) at a higher temperature (384 K) and its effect on the electrical properties to invoke the necessity for a ferroelectric-to-paraelectric transition at a lower temperature. While a high-temperature transition may exist—at least based on their published results (for detailed discussion, see section 8 below)—in our original article and the related work, we did not study the possibility of higher-temperature phase transitions of this material and their structure(s) and properties because they are irrelevant to our interest in the dynamic properties of the crystals of guanidinium nitrate as an actuating material. It appears that MS and AK continue to ignore that fact. As clearly shown in the DSC in Figure 1d, in our work we focused on the behavior of the material below 310 K. As it is also explained in detail above, **our interest in this material is limited to its performance as an actuating material that can operate by structure switching close to room temperature**. We would also like to refrain from commenting on the accuracy of the results of MS and AK related to the details of the other phase transitions of this material, although we think that they warrant becoming the subject of further investigations for verification of their results from the perspective of basic research. More importantly, MS and AK assume that polycrystalline and single crystals exhibit identical polarization behavior under an electric field, disregarding the widely accepted fact that the electric polarization of the ferroelectrics is strongly affected by the form of the material (i.e., powder, single crystal, or film), the method of growth, and the quality of the crystals. Polycrystalline powder samples do not have uniform orientation of their crystal lattices, in addition to heterogeneity brought about by a range of particle sizes, grain boundaries, and porosity, all of which might lead to cancellation of the dipole moment close to room temperature.

We were also astonished to find—after a careful inspection of the details of the new experiments reported for a single crystal in the revised comment by MS and AK—*that the results and the frequency, voltage, temperature, and experimental conditions used are remarkably similar to those in their original comment*. Their measurements use different frequency and voltage ranges than ours. The details of their experimental testing of a single crystal, its phase identity, and confirmation of each form prior to electrical testing are not reported and remain unclear. These inconsistencies invalidate their conclusions. Figure 2a in the revised comment by MS and AK shows the *P–E* loop of both a powder sample and a single crystal <u>only</u> of form III (referred to as form I in our article). As before, there is no mention of the frequency used in their experiment; however, their earlier works[20,21] regularly report *P–E* loops recorded at a single frequency of 50 Hz. Based on these results, MS and AK unjustifiably dismiss the possibility of spontaneous polarization in our samples of single crystals of forms I and II guanidinium nitrate, while it is well known that the measurements of the *P–E* and *C–V* hysteresis depend on the frequency, temperature, and voltage. Moreover, the results are remarkably sensitive to the structural and phase impurities of the crystals, the method of growth, the quality of the crystals, as well as the direction of the applied electric field, all of which are not available from their revised comment.

Furthermore, in Figure 2b of their comment, MS and AK report *C–V* plots of phase III (form I in our notation) and phase II (referred to as form II by us). Contrary to their claim of frequency-dependence in their measurements, it is evident from their Figure 2b and the respective caption ("*Capacitance–electric field dependence measured on a single crystal in phases II and III at 285 K.*") that these measurements are *not* frequency-dependent. Although we were not able to identify the specific frequency at which these measurements were conducted, it is apparent that their *C–V* plots have not been recorded during a complete voltage sweep, which involves sweeping the voltage up and down (positive to negative, followed by back to positive) with a defined holding period. A hysteresis would be observed if such voltage sweeps were performed, and the magnitude of the hysteresis is known to be frequency-dependent. It is evident that **these measurements do not align with the frequency-dependent *C–V* measurements described in our article, as they were performed at different voltage ranges at an unknown frequency, and perhaps most importantly, they do not consist of a double sweep with a defined holding time.** Interestingly, MS and AK claim to have used the same capacitor size as in our experiments; however, they did not find it necessary to provide any information on the methodology employed or details regarding the construction of the individual capacitors for each confirmed crystal phase. This critically important information further calls into question their conclusions. Their claims are also surprising, given that in their previous work to which they refer in their comment, and a conclusion of which is copied below (Figure 19), they were unable to perform dielectric measurements on single crystals of guanidinium nitrate due to their small size: [9]

> As the obtained single crystals of guanidinium nitrate were too small for dielectric measurements, we had to perform these measurements for polycrystalline samples.

**Figure 19**. A copy of the text published in ref.[9] Adapted with permission from ref. 9. Copyright 1992, Elsevier.

**Comment:** *In section Electrical properties, Karothu et al. describe experiments, where they placed the GN crystals between metal plates made of Ag and Cu; the choice of different electrodes is unfortunate because it usually causes asymmetric results with respect to the zero field[11]. Furthermore, metal plates do not guarantee good contacts with the crystal surfaces, but spurious capacitors can be formed, which can distort the results. Therefore, to perform reliable dielectric measurements, the electrodes should be deposited on the crystal surfaces by sputtering, evaporation, or painting. Besides, a dry atmosphere of measurements is mandatory, since the water adsorbed on the crystal surfaces can dramatically affect the results. None of these requirements was met by Karothu et al. Consequently, they obtained unrealistic high capacitance values for their capacitors containing GN in phases III and II, incompatible with the size of the crystals described in the paper nor with the true values of electric permittivity. The capacitance values of about 6.2 nF in phase III and 55 nF in phase II (their Fig. 4c) juxtaposed with the dimensions of the crystals (their Fig. 1, Supplementary Fig. 10 and Table 5) indicate that the electric permittivity of GN at 200 kHz frequency would have to be in the range of tens of thousands to hundreds of thousands at least, while the true $\varepsilon$' value is 4.4–5.0 only (our Fig. 1c). We confirmed such relatively low $\varepsilon$' values, typical for non-ferroelectric dielectrics, also for the single-crystal samples. Noteworthy, the difference in electric permittivity between GN phases III and II amounts to mere 10%, while two orders of* *magnitude higher change is claimed by Karothu et al. Their huge capacitance reaching thousands nF (cf. their Supplementary Table 6), can originate from the high conductivity of their samples exposed to the humid air or from other experimental errors. In Fig. 2c we compare our conductance data measured for single crystal of GN with those reported in their paper. The difference is incredible as it reaches ten orders of magnitude. The 1 kHz conductivity of GN in phase II derived from our data amounts to $2.6 \cdot 10^{-9}$ S/m, which is consistent with the dielectric character of the material, while the value of 17.6 S/m was estimated from the Karothu et al. data (the dimensions of the sample were not specified and therefore we assumed the capacitor thickness 0.1 mm and the electrode area 3 mm$^2$ for this estimation on the basis of the size of crystals presented in their paper). Such a high electric conductivity measured by Karothu et al.$^1$ is characteristic of semiconductors rather than of insulators. Moreover, it is well known that the conductance of a capacitor filled with dielectric material increases with the frequency of electric field, which is consistent with our measurements plotted in Fig. 2c, while Karothu et al$^1$ reported the opposite relationship. Their explanation of the capacitance/conductance decrease by switching off the dipolar and ionic polarization contributions in the frequency range from 1 to 200 kHz is unjustified, because in solid dielectrics the dipolar relaxation occurs in the MHz–GHz frequency range and the resonant frequencies of ionic contributions fall into the infrared region. In our studies, the low conductivity of GN is confirmed by the very small dielectric losses in both phases III and II around room temperature (our Fig. 1c). The parameters of the capacitors, such as capacitance and conductance presented by Karothu et al., instead of electric permittivity, dielectric loss, and conductivity, are inadequate for characterizing the material properties.*

**Response to the comment:** We find it absurd and completely unacceptable that, on several occasions in their revised comment, MS and AK continue to insist that we have used an asymmetric capacitor configuration and that we claimed the semiconducting properties. It also appears that MS and AK, intentionally or not, did not consider the entirety of the detailed electrical measurements presented in our article. For the frequency-dependent *C–V* and *G–V* measurements, we used the four terminal-pair configuration impedance measurements of the Agilent B 1505A curve tracer and manual probes (a schematic of the configuration is provided in Figure 4a of our article[8]). In this experiment, two **symmetric silver electrodes** were deposited on the top and bottom surfaces of single crystals of forms I and II by squeezing the crystal in between two silver thin films in a vertical configuration. The crystal was then placed on a copper plate to be able to access the crystal bottom electrode, as illustrated in Fig. 4b in ref. 8. Additionally, current-voltage measurements (leakage current) were performed on single crystals of forms I and II of guanidinium nitrate by depositing two separate silver electrodes on the top surface of the crystal (lateral configuration), as shown in Fig. 5a in ref.[8]. Similar testing is widely used for the characterization of organic electronic devices (for one such example, see ref.[22]). The main objective of our electrical testing was to investigate how the changes in the lattice (from form I to form II) affect their electrical characteristics (dipole strength, ions, and transport properties).

In their comment, MS and AK criticize the conductance and capacitance values obtained in our study, questioning the quality of the silver contact with the crystal (interfacial effect) and the effect of the humid air in our measurements. However, it appears that they have overlooked these details and their (possible) effects, which have already been addressed in our article[8] as well as in our previous response: "*The four types of polarization are electronic, ionic, dipolar, and* *interfacial*, *and all of them can effectively add to the capacitance and conductance values at low-frequency range.* **At high frequencies, the interfacial, dipolar, and ionic polarization contributions become insignificant, leaving only the contribution from the electronic polarization relevant.** *This frequency dependence can be clearly seen in both phases of the crystal. As shown in Supplementary Table 6, the capacitance/conductance decreases as the frequency increases from 1 to 200 kHz. This is due to the fact that at higher frequencies the temporal response of the* **interfacial dipoles** *is reduced, leading to a lack of rearrangement in the alternating fields direction."* Furthermore, we write: *"Additionally, the quality of the interface between the silver and the crystal can contribute to the observed values, most probably by physical adsorption of a thin film of water on the crystal surface which can occur with testing in air."* In any given material, a number of these polarization contributions are expected to be active at different frequencies. As explained above, further studies by *C–V* and *G–V* measurements at frequencies above 200 kHz (500 kHz – 1 MHz) were carried out by us and reported in our original article[8] for a crystal of form I, and a capacitance value of ~46 pF at 1 MHz was obtained.

The multi-frequency capacitance unit module, utilized in our testing, performs the impedance measurement and returns the *C–V* and *G–V* curves (i.e., the impedance *vs.* frequency characteristics of a capacitive load). The *C–V* and *G–V* curves provide *direct* information about the electric field inside the material and are not directly related to the variation of carriers in the crystal (i.e., only the AC ionic conductivity can be recorded but not the DC transport properties of the material). Therefore, to extract the electrical conductivity of the two forms, we performed current–voltage measurements, as shown with the results in Figure 5 in our article.[8] The *I–V* measurements are very well established in organic electronics research and are performed to deduce the transport properties of the respective materials. For the purpose of *I–V* testing in our study, two separate silver electrodes were deposited on the top surface of the crystal, and eight single crystals of both form I and form II were used in the measurements. During the experiments, we could clearly detect the presence or absence of water on the surface; when condensation occurred on the crystal surface, as expected, poor *I–V* characteristics were obtained because water molecules increase the energy barrier for transportation of charge carriers and reduce their mobility. This is in contrast with the comments of MS and AK, who state[7] that conductance increases in humid air. In fact, the *electronic conductivity decreases* with increasing relative humidity based on our *I–V* results, whereas the *ionic conductivity increases* with increasing relative humidity in a low-frequency range. These two separate measurements have already been discussed and well addressed in our article.[8] Panels a and b in Figure 20 below show the linear scale of the *I–V* characteristics of forms I and II of guanidinium nitrate (the logarithmic scale and current density calculations are available from the original article[8]) and the electrical conductivity $\sigma = L/RA$, where $L$ is the distance between the silver electrodes, $A$ is the area (estimated as the crystal thickness multiplied by the crystal width), and $R$ is the resistance calculated as $\Delta V/\Delta I = 1/\text{slope}$.

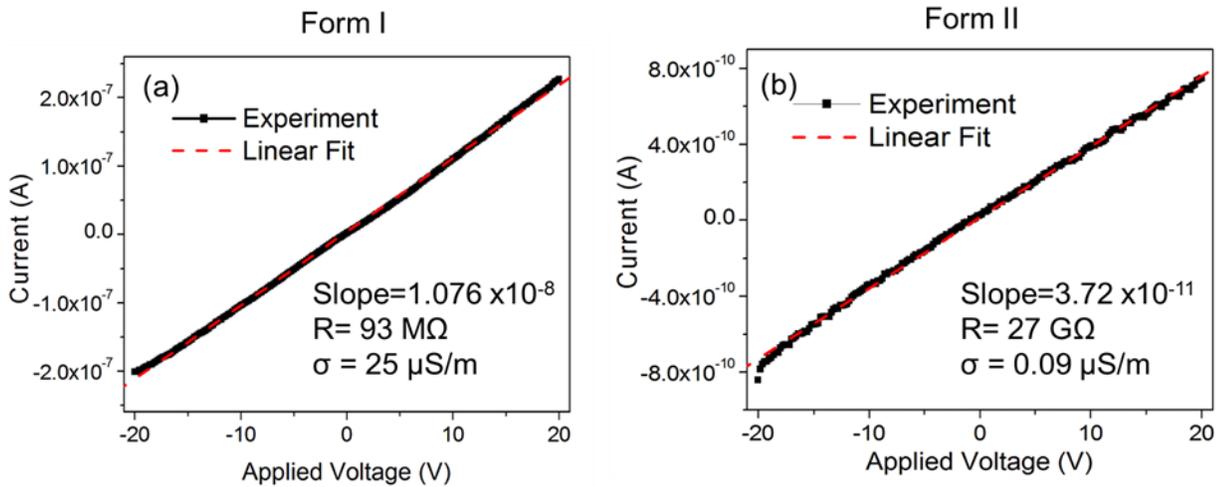

**Figure 20**. The *I—V* (current—voltage) characteristics of crystals of guanidinium nitrate in form I and form II. (a) Ag/(form I crystal)/Ag, (b) Ag/(form II crystal)/Ag.

In Figure 2c of their revised comment, MS and AK show a comparison between the frequency-dependent conductance of a single crystal and our data on ionic conductivity. Once again, there is a lack of information regarding the phase identity of the crystal under study and the specifics of the experimental procedure used. For instance, there is no mention of the type of voltage source (AC or DC voltage). Nevertheless, based on our data and the information presented in the same curve, we assume that the sample used by MS and AK corresponds to a form II crystal. Intentionally or not, it appears that MS and AK did not consider the entirety of the detailed electrical measurements presented in our article that were also highlighted and explained in our previous response. Based on our *I–V* measurements (Figure 5 in our article[8] and Figure 20 above), the electrical conductivity of form I is 25 µS/m and that of form II is 0.09 µS/m. These values are small and far from the typical semiconductor range. Furthermore, in our article, as it was already discussed above, we have not made any claims about the material being semiconductive, and it is perplexing that MS and AK have directly attributed to us explicit claims that are not present in our report or have ever been a subject of our research.

Having said this, it is also clear that the details provided in our response and in our original article regarding the differences between the *G–V* measurements and the *I–V* measurements (not included in their experiments) have been overlooked by MS and AK. Namely, the higher values of *C* and *G* at lower and moderate frequencies are the result of four types of polarization: electronic, ionic, dipolar, and interfacial. All of these can effectively add to the capacitance and conductance values in the low-frequency range. At high frequencies, the interfacial, dipolar, and ionic polarization contributions become insignificant, leaving only the contribution from the electronic polarization relevant. Furthermore, the metal/crystal interface is important, as it has been discussed in detail both in our article and in our previous response. Hence, further studies by *C–V* and *G–V* measurements at frequencies above 200 kHz (500 kHz – 1 MHz) were carried out by us, and are reported in our article[8] for a crystal of form I, where a value for the capacitance of ~46 pF at 1 MHz was obtained.

By comparing the electrical performance of crystals of form I and form II from the *G–V* and the *I–V* measurements, we have already established that during the transition from form I to form II, the inevitable accumulation of crystal defects may be the source of inner local stress (although this would not prevent cyclability of the macroscopic mechanical response, as shown in Figure 2e in our article[8]), and they can be responsible for the reduction in current density seen in form II (Figure 5b). This is also consistent with the higher conductance values obtained for the crystal of form II, where the defects can act as fixed charges that add up to the conductance values. As mentioned above, the main objective of our electrical measurements on this material was to establish how the change of the lattice from form I to form II affects the material's electrical characteristics, and that was done by comparing the performances of the two crystal forms. This point has already been discussed in the original article:[8] "*As can be seen, the form II crystal shows seven-fold higher capacitance compared to form I. Similarly, form II exhibits 35-fold increase in conductance compared to form I at the same frequency (Fig. 4d).*" Also, we write: "*The current density (J = current/area) calculated for the two crystal phases was J = 0.28 mA cm$^{-2}$ and 0.8 µA cm$^{-2}$ for form I and form II, respectively.*" These types of measurements, aimed at comparing the electrical performance of two crystal forms of the same material, are not uncommon in research on organic electronic systems.

**Comment:** *Their paper and its Supplementary Info. contain many other serious errors. For example, their structural mechanism of the phase transition based on the rotations of hydrogen-bonded rings built of three ionic pairs (their Figs. 1g-i, Supplementary Figs. 6 and 11) is unrealistic because of the high energy required for breaking 12 hydrogen bonds and angular momentum connected to such collective rotations. The calculations of energy associated with the rotation of supramolecular rings and a continuous doubling of the unit cell (their Supplementary Fig. 11a-d) cannot be reconciled with the first-order character of the phase transition. Previous $^1$H NMR results$^{14}$ show that less energetic reorientations of guanidinium cations around their pseudo C3 axes are activated only in GN phase II above 300 K. Such reorientations require breaking and subsequent formation of six N–H···O hydrogen bonds. Thus, the transition mechanism involving concerted rotations of rings built of six ions tied within the layer with twelve H-bonds, at a much lower temperature in addition, is questionable.*

**Response to the comment**: In a rather vague statement, MS and AK go on to refer to "many other serious errors" concerning the supporting data in our article before focusing singularly on the phase transition mechanism and specifically the insights gained from our dispersion-corrected density functional theory (DFT-D) calculations ("*their structural mechanism of the phase transition based on the rotations of hydrogen-bonded rings built of three ionic pairs (their Figs. 1g–i, Supplementary Figs. 6 and 11) is unrealistic because of the high energy required for breaking 12 hydrogen bonds and angular momentum connected to such collective rotations*"). At the outset, it is important to note that the mechanism of the phase transition of guanidinium nitrate was not the central focus of our paper. The DFT-D calculations were performed in order to gain a better understanding of the rapid martensitic phase transition between forms I and II of guanidinium nitrate on energetic grounds by focusing on two aspects of the phase transition: molecular rotations of the ions and the strain build-up of the crystal. The above comment by MS and AK attempts to impose conclusions regarding the phase transition that are inconsistent with our own theoretical data and indeed with the numerous examples of martensitic phase transitions in molecular crystals previously reported in the literature.[23] It is unfortunate that MS and AK make these assertions without providing any theoretical data of their own that may shed light on the phase transition.

Additionally, it is well known that martensitic phase transitions in molecular crystals can proceed via the collective movement of molecular units, *assisted specifically by the rotation of molecular units within the crystal*.[23]

In the case of the two polymorphs of guanidinium nitrate studied in our work, the main packing difference between the two phases is the stacking of the molecular ions **between the layers.** As such, a phase transition based on molecular rotations is not only consistent with our crystallographic observations but also with numerous literature examples of martensitic phase transitions that proceed via such collective rotations.[23] MS and AK also claim that hydrogen bonds must be broken during the phase transition, and hence there is a "high energy barrier" for the phase transition. The fact that the crystals of forms I and II are equienergetic means that whatever local changes happen to the hydrogen bond network during the phase transition (a subject outside the scope of our theoretical work), the static DFT-D energy of forms I and II remains practically the same. Thus, the claim of MS and AK that the phase transition involves "high energy" is without foundation since MS and AK have not provided any computational data of their own to corroborate their assertions. By contrast, the state-of-the-art DFT-D calculations disclosed in our article[8] provide valuable insights into the energetics of the phase transition using well-established literature procedures and codes. In conclusion, our DFT-D calculations use an idealized bilayer model to probe the rotational energy barrier for cooperative motion of the hydrogen-bonded ions in this material (Supplementary Figures 11a, b in ref. 8). This clearly shows that rotation of the hexagonally hydrogen-bonded ions is energetically favorable to happen. Separate calculations shown in Supplementary Figures 11c and 11d looked at the energy barrier for applying crystal strain as we transformed from form I to form II. Our calculations reveal that this process is energetically costly, and we have quantified this using DFT-D methods at each strain step. It is important to note that these are *separate DFT-D calculations* designed to probe the energetic penalty involved in specific parts of the martensitic phase transition, a point that MS and AK conveniently ignored in their comment. We cannot comment on the combination of both processes (rotations and strain) inside the same crystal and how they play a role in the phase transition, as a detailed theoretical study devoted to the phase transition—such as those typically seen in nudged elastic band (NEB) calculations—was deemed outside the scope of our work. Nevertheless, our calculations provide a clear picture of the key factors at work in the phase transition since both the rotations of the hydrogen-bonded ions and the strain build-up in the crystal are necessary for the observation of the martensitic phase transition as well as the bulk properties observed for this dynamic organic crystal actuator. A key finding from our computational studies is that *"the ion stacking in the two phases is equienergetic, favoring rapid switching between them"*. **The computational data therefore explain why this material rapidly switches between two polymorphs,** and the rotational dynamics we investigated in our calculations are consistent with both the crystallographic observations of differences in the ion stackings as well as prior literature on the key drivers for martensitic phase transitions in molecular crystals.

**8) On the (lack of) justification of using the higher-temperature transition between phases II and III to discredit the evidence of the electrical properties of phases I and II**

In their comment, MS and AK take their earlier crystallographic results on form III and the transition from form II to form III as an argument to question our interpretation of our results on phase I properties (all in our notation). Specifically, MS and AK comment that "*GN also undergoes a second-order transition at about 384 K. This is of primary importance, because the* *high-temperature phase I is centrosymmetric, of space group $R\bar{3}m$.[7]*", and in the DSC in their Figure 1b, they even add this space group as the definitive symmetry of the high-temperature phase, form III (their Figure 1b is copied as Figure 16 above). Although, as we mentioned above, the high-temperature transition has never been a subject of our study, which focuses on the transition between forms I and II, a careful inspection of the results of MS and AK's earlier publications, motivated by their comment, showed that the high-temperature phase transition as well as the related phases are not as well defined or understood as they would like to present it.

In fact, their published results indicate ambiguities in both the existence and the structure of form III (corresponding to phase I in their later notations). In their previous works, their form II (corresponding to our form II) has been inconsistently assigned to both space groups *C*2 and *C*2/*m* in several of their published articles (for example, see Figures 21 and 22, copied from refs.[14,15]). While both of these space groups are monoclinic, they differ in centricity; one of them is non-centrosymmetric (*C*2), while the other is centrosymmetric (*C*2/*m*):

> antiparallel in monoclinic (*C*2/*m*) phase II; and parallel in
> monoclinic (*Cm*) phase III. The collapse of voids in phase IV is

**Figure 21.** A copy of the text published in ref.[15] Form II is claimed to have symmetry *C*2/*m*. Adapted with permission from ref. 15. Copyright 2011, Royal Society of Chemistry.

> sulting in a nearly 50% elongation of the crystals. The
> symmetries of the low-temperature GN1 phase (*Cm*),
> intermediate GN2 phase (*C*2) and high-temperature
> GN3 phase (*C*2/*m*) were determined by X-rays. As the

**Figure 22.** A copy of the text published in ref.[14] Adapted with permission from ref. 14. Form II is claimed to have symmetry *C*2. Copyright 2004, Elsevier.

Table 1 below, which summarizes the space groups published by MS and AK before, reflects more contradictory and/or missing data in the space groups that were used by these two authors to model the phases and the phase transitions of guanidinium nitrate:

**Table S1**. Space groups of the phases of guanidinium nitrate and the respective experimental conditions available from the earlier works published by MS and/or AK

| Form (our/their notation) | Temperature (K) | Space group | Reference and comment(s) |
|---|---|---|---|
| I (form III) | Room temperature (293–303) | *Cm* | [11] |
| II (form II) | Room temperature 283–303 | *C*2 | [12] |
| II (form II) | medium temperature range | *C*2/*m* | [15] |
| II (form II) | at 360000.0 kPa, 295 K | *C*2/*m* | [15] |
| III (form I) | 391 | *C*2/*m* | [12] Several constraints and restrains have been used. |

| III (form I) | 384 | $R\bar{3}m$ | Taken from a comment by MS and AK filed on a preprint server. **No structure has been deposited within the CSD for form III with this space group.** The reference cited in the comment paper corresponds to deuterated guanidinium nitrate. The unit cell parameters were obtained from powder X-ray diffraction and neutron diffraction. |
| III (form I) (deuterated) | 395 | $R\bar{3}m$ | Taken from Ref. 14. **No structure has been deposited within the CSD for form III with this space group**. The unit cell parameters were obtained from powder X-ray diffraction and neutron diffraction. |

The form III claimed by MS and AK in their comment form I in their later notation) was described in their article given here as ref. 12, where a phase transition between form II and another form, form III, was attributed to an anomaly observed in the thermal behavior of guanidinium nitrate around 391 K. Diffraction data were collected by using single crystal X-ray diffraction around that temperature, and very small changes in the unit cell parameters were determined (Figure 23).[12] The structure of the higher-temperature phase, form III (their form I), was assigned the space group ($C2/m$).

| GN2 | 292 | 12.545(5) | 7.303(4) | 7.476(4) | 124.93(5) | 561.55 |
| | 335 | 12.618(17) | 7.280(5) | 7.499(10) | 123.96(15) | 571.36 |
| | 375 | 12.590(21) | 7.286(12) | 7.566(12) | 123.73(18) | 577.17 |
| GN3 | 384 | 12.610(24) | 7.285(6) | 7.579(14) | 123.70(21) | 579.20 |

**Figure 23.** A copy of the text published in ref.[12] Adapted with permission from ref. 12. Copyright 1992, Elsevier.

However, the poor quality of the data and the low accuracy of the refined basic structural parameters of the claimed form III (their form I) by AK and MS and the similarity of its unit cell parameters with form II (their form II) point out ambuiguities in the occurrence and the real nature of this phase transiton, which has been postulated based on thermal data and powder X-ray diffraction data. At the time of preparation of this manuscript, **the actual structure of form III remains unclear without firm experimental evidence based on good-quality single crystal diffraction analysis,** and it probably warrants further studies to be considered a real phase transition that not an artifact of the process of sample preparation or other factors. Ironically, similar conclusions were reached by the authors (Figure 24):[12]

system proved unsuccessful. The lower quality of the GN2 and GN3 samples is reflected in a lower accuracy of the determination of the unit-cell dimensions of the GN2 and GN3 phases (see Table 1), in some ambiguity of the space-group assignments, in high *R*-factors (Table 2), and in

**Figure 24.** A copy of the text published in ref.[12] Adapted with permission from ref. 12. Copyright 1992, Elsevier.

We further note that the alleged form III (form I in their later notation) was once reported by these authors in space group $C2/m$[12] and once in $R\bar{3}m$[14] (the latter space group was also claimed in their comment), **but no crystal structure obtained from a single crystalline sample has been deposited in the CSD with the latter space group**. Ref. 7 cited in MS and AK's comment corresponds to a deuterated sample of guanidinium nitrate[14] analyzed by powder X-ray diffraction and neutron diffraction. In the same article,[14] a higher symmetry was suggested for phase III based on conoscopic patterns obtained from *prismatic* plate-shaped guanidinium nitrate crystals. However, in their earlier article (Figure 25), a higher symmetry was also suggested for form II (their form II):[12]

> refined on the data measured for the GN2 phase at 325 K converged to $a = 12.6403(3)$, $b = 7.30184(8)$, $c = 7.4618(8)$ Å, $\beta = 123.857(2)°$, corresponding to the pseudo-hexagonal cell of $a = 7.299$, $c = 18.590$ Å, $\alpha = 90.45$, $\beta = 89.55$ and $\gamma = 119.57°$.

**Figure 25.** A copy of the text published in ref.[12] Adapted with permission from ref. 12. Copyright 1992, Elsevier.

Given the insistence of MS and AK to resort to the high-temperature phase (phase III), advanced in order to discredit the results of our measurements on gaunidinium nitrate in the lower temperature phases (form I), we performed additional experiments to verify independently some of their claims regarding the high-temperature phase transition. In our hands, during single crystal X-ray measurements that were aimed to explore the purported phase transition from form II (their form II) to form III (their form I), we observed that the sample undergoes sublimation at high temperatures (380 to 395 K). Fitting the minute changes into reliable models of structures that are related by a phase transition proved to be extremely unreliable due to the sublimation. In fact, the same conclusion was also reached previously by MS and AK (Figures 26 and 27),[12] however, it was not mentioned in their comment:

> 292 K; however, at 393 K the GN3 crystal sublimed slowly and the intensity of control reflections dropped by 15% over the span of the data collection — appropriate corrections were applied to the

**Figure 26**. A copy of the text published in ref.[12] Adapted with permission from ref. 12. Copyright 1992, Elsevier.

> The GN3 structure was refined in space group *C*2/*m*, while all attempts to refine it in a hexagonal system proved unsuccessful. The lower quality of the GN2 and GN3 samples is reflected in a lower accuracy of the determination of the unit-cell dimensions of the GN2 and GN3 phases (see Table 1), in some ambiguity of the space-group assignments, in high *R*-factors (Table 2), and in a lower precision of the atomic positions in the refined models of the GN2 and GN3 structures (Tables 4 and 5). This particularly concerns the GN3 phase, where the measurements and calculations were hampered by sublimation of the sample

**Figure 27**. A copy of the text published in ref.[12] Adapted with permission from ref. 12. Copyright 1992, Elsevier.

Although this has not been the focus of our work, to verify the earlier results of MS and AK on the high-temperature phase transition, we further recorded DSC on both a single crystal and a lightly ground powder of guaninium nitrate at higher temperatures than our previous measurements. Our results confirmed the occurrence of a very small anomaly at higher temperatures (Figure 28), which is related to a much lower enthalpy than the low-temperature phase transition. However, our results also indicated that this anomaly is accompanied by significant sublimation of the sample on heating, and the comparison of the appearance of the DSC curve indicated that it may be due to what is typically observed with order-disorder transitions. And while we agree that this transition requires further investigation, given the above uncertainties with the structure determination, we would refrain from making any claims as to its nature, and even less so from using this transition between form II and another elusive phase (III) to draw conclusions on the low-temperature phase transition between forms I and II.

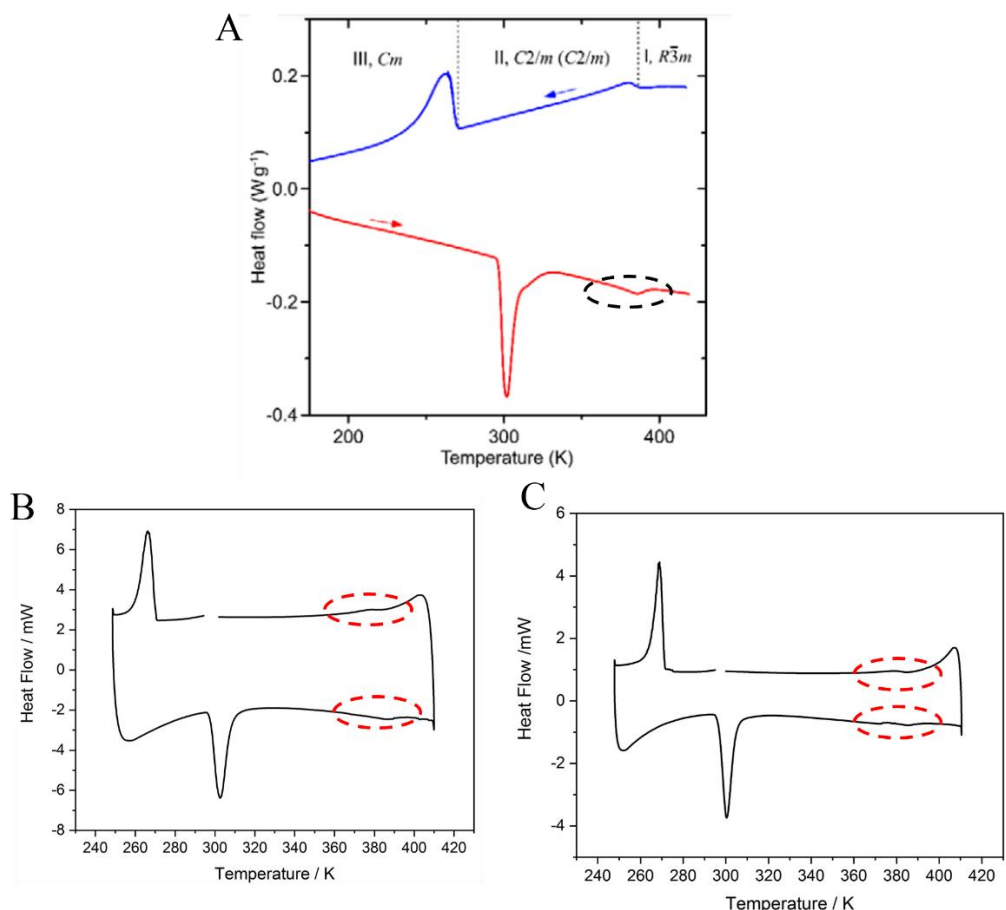

**Figure 28**. Differential Scanning Calorimetry (DSC) on guanidinium nitrate. (A) Result of MS and AK copied from their comment. (B,C) Our DSC curves recorded from a single crystal (B) and a lightly ground sample (C). The region marked with broken circle highlights a high-temperature anomaly that has been previously attributed by MS and AK to a phase transition between forms II and III, with a poorly characterized structure in form III.

**Comment**: *In conclusion, the structural determinations of GN crystals as a function of temperature and pressure, phase transitions, giant deformation, its detailed structural mechanism, molecular dynamics, and dielectric properties of GN were reported before[2-7,9,14]. Undoubtedly, this interesting compound of intriguing properties deserves further studies. However, the citation of previously published results is a generally accepted ethical standard, indispensable for the progress in science. In the 1990s, we did not observe the semiconductivity, ferroelectricity, or fatigue resistance of the giant deformation in GN crystals and our extended measurements presented above do not confirm such properties, either. In the light of our results, the properties of GN crystals are unsuitable for the applications in light-weight capacitors, ferroelectric tunnel junctions, thermistors and multiple-stroke actuators, as proposed by Karothu et al.[1]*

**Response to the comment**: We remain steadfast in our commitment to maintaining the highest scientific standards and fostering an environment of open and honest discourse. An open debate and the freedom to criticize published work are some of the underlying principles of the scientific method. Reasonable arguments, brought to the attention of the public in a follow-up communication by experts who are knowledgeable in the field and could provide a reassessment of data, an interpretation, and conclusions, are always welcome and encouraged. For a peer-reviewed publication, this is a process that requires a strong and well-rounded justification, supported by relevant literature and/or additional experiments, as a means to demonstrate the lack of validity of a concept or conclusion that has been filed and exists as public knowledge in the scientific literature.

However, comments on primary publications are only justified when they are made public with the intention of correcting the scientific literature for the sake of promoting the scientific truth on a certain subject. Comments do *not* make sense when they are published for the purpose of promoting one's own work, with an intention to force other authors who work on the same subject to cite previously published work even that they do not consider it directly relevant—particulary when it is not well supported by experimental evidence and many important aspects remain questionable, such as this case—or as a means for discrediting another's work without a strong scientific basis. Such attempts, which can hardly be accepted as an intention to uncover the ground truth and can easily be spotted when authors frequently write comments,[1–7] point out other factors that drive that practice, such as a feeling of a lack of attention or underappreciation of someone's work to one's work by their peers, inconspicuous act(s) of retaliation, or simply an intention to garner more citations.

## References

(1) Szafrański, M. Comment on "1,4-Diazabicyclo[2.2.2]octane-based disalts showing non-centrosymmetric structures and phase transition behaviors" by X.-B. Han, P. Hu, C. Shi and W. Zhang, CrystEngComm, 2016, 18, 1563†. *CrystEngComm*, **19**, 179–182 (2017).

(2) Szafrański, M. & Katrusiak, A. Comment on "Improper molecular ferroelectrics with simultaneous ultrahigh pyroelectricity and figures of merit" by Li et al. *Sci. Adv*. **8**, eabi6220 (2022).

(3) Szafrański, M. Comment on "Phase transitions, screening and dielectric response of $CsPbBr_3$" by Š. Svirskas, S. Balčiūnas, M. Šimėnas, G. Usevičius, M. Kinka, M. Velička, D. Kubicki, M. E. Castillo, A. Karabanov, V. V. Shvartsman, M. R. Soares, V. Šablinskas, A. N. Salak, D. C. Lupascu and J. Banys, J. Mater. Chem. A, 2020, 8, 14015. *J. Mater. Chem. A*, **9**, 11450—11452 (2021).

(4) Szafrański, M. Comment on Unprecedented 30 K hysteresis across switchable dielectric and magnetic properties in a bright luminescent organic—inorganic halide $(CH_6N_3)_2MnCl_4$ by A. Sen, D. Swain, T. N. Guru Row and A. Sundaresan, J. Mater. Chem. C, 2019, 7, 4838. *J. Mater. Chem. C*, **8**, 2594—2596 (2020).

(5) Szafrański, M. Comment on "Ferroelectricity in bis(imidazolium) L-tartrate", *Angew. Chem. Int. Ed*. **52**, 7076—7078 (2013).

(6) Szafrański, M. Comment on the "Phase transition mechanism in diglycine methanesulfonate",

*Chem. Asian J*. **9**, 3342—3343 (2014).

(7) Szafrański, M. & Katrusiak, A. On the giant deformation and ferroelectricity of guanidinium nitrate. (both the original comment and the revised comment are available from this link) https://arxiv.org/abs/2301.13481

(8) Karothu, D. P. et al. Exceptionally high work density of a ferroelectric dynamic organic crystal around room temperature. *Nat. Comm*. **13**, 2823 (2022).

(9) Szafrański, M. Unusually strong deformation of guanidinium nitrate crystal at the solid-solid phase transition. *Solid State Commun*. **84**, 1051–1054 (1992).

(10) Szafranski, M., Czarnecki, P., Dollhopf, W., Hohne, G. W. H., Brackenhofer, G. & Nawrocik, W. Investigation of phase transitions in guanidinium nitrate crystals. *J. Phys.: Condens. Matter* **5**, 7425–7434 (1993).

(11) Katrusiak, A. & Szafrański, M. Guanidinium nitrate. *Acta Cryst. C* **50**, 1161–1163 (1994).

(12) Katrusiak, A. & Szafrański, M. Structural phase transitions in guanidinium nitrate. *J. Mol. Struct.* **378**, 205–223 (1996).

(13) Szafrański, M. Influence of molecular and lattice vibrations on the stability of layered crystal structure of guanidinium nitrate. *Phys. Stat. Solidi (b)* **201**, 343–354 (1997).

(14) Szafrański, M. & Katrusiak, A. Temperature-induced H-site centring in NH--O hydrogen-bonds of guanidinium nitrate by neutron diffraction. *Chem. Phys. Lett*. **391**, 267—272 (2004).

(15) Katrusiak, A., Szafrański, M. & Podsiadło, M. Pressure-induced collapse of guanidinium nitrate N—H···O bonded honeycomb layers into a 3-D pattern with varied H-acceptor capacity. *Chem. Commun*. **47**, 2107—2109 (2011).

(16) Rahman, M., Kundu, S. K., Chaudhuri, B. K. & Yoshizawa, A. Frequency-dependent capacitance voltage hysteresis in ferroelectric liquid crystals: An effect of the frequency dependence of dielectric biaxiality. *J. Appl. Phys*. **98**, 024114 (2005).

(17) Mao, D., Quevedo-Lopez, M. A., Stiegler, H., Gnade, B. E. & Alshareef, H. N. Optimization of poly(vinylidene fluoride-trifluoroethylene) films as non-volatile memory for flexible electronics. *Org. Electron.* **11**, 925—932 (2010).

(18) Huadong L. & Subramanyam, G. Capacitance of thin-film ferroelectrics under different drive signals. *IEEE Trans. Ultrason. Ferroelectr. Freq. Control* **56**, 1861—1867 (2009).

(19) Misirlioglu, B. & Yildiz, M. Carrier accumulation near electrodes in ferroelectric films due to polarization boundary conditions. *J. Appl. Phys*. **116**, 024102 (2014).

(20) Szafrański, M. Ferroelectricity in the guanidinium compound [C(NH$_2$)$_3$]$_4$Cl$_2$SO$_4$: Synthesis and characterization. *Phys. Rev. B* **72**, 054122 (2005).

(21) Szafrański, M. Simple guanidinium salts revisited: room—temperature ferroelectricity in hydrogen-bonded supramolecular structures. *J. Phys. Chem. B* **115**, 8755—8762 (2011).

(22) Naber, R. C. G., Asadi, K., Blom, P. W. M., de Leeuw, D. M. & de Boer, B. Organic nonvolatile memory devices based on ferroelectricity. *Adv. Mater*. **22**, 933—945 (2010).

(23) Park, S. K. & Diao, Y. Martensitic transition in molecular crystals for dynamic functional materials. *Chem. Soc. Rev*. **49**, 8287—8314 (2020),

**Data availability** All data obtained and analysed during this study are included in the article.